\documentclass[aps,preprint,floatfix,nofootinbib,showpacs]{revtex4}
\usepackage{xcolor}
\usepackage{graphicx,color,bm}
\usepackage{amsmath,amssymb}
\usepackage{hyperref}
\usepackage{epstopdf}
\usepackage{ulem}
\usepackage{array}
\usepackage{verbatim}
\usepackage[utf8]{inputenc}
\renewcommand{\baselinestretch}{1.1}

\usepackage{rotating}

\newcommand{\mc}{\mathcal}
\newcommand{\p}{\partial}
\newcommand{\f}{\frac}
\newcommand{\dd}{\text{d}}

\newcommand{\be}{\begin{equation}}
\newcommand{\ee}{\end{equation}}
\newcommand{\ben}{\begin{equation*}}
\newcommand{\een}{\end{equation*}}
\newcommand{\bea}{\begin{eqnarray}}
\newcommand{\eea}{\end{eqnarray}}
\newcommand{\bean}{\begin{eqnarray*}}
\newcommand{\eean}{\end{eqnarray*}}

\newcommand{\kpc}{\ensuremath{\,\mathrm{kpc}}}
\newcommand{\Mpc}{\ensuremath{\,\mathrm{Mpc}}}
\newcommand{\ev}{\ensuremath{\,\mathrm{eV}}}

\newcolumntype{K}[1]{>{\centering\arraybackslash}m{#1}}

\begin{document}

{\small
\begin{flushright}
NCTS-PH/1719\\
\end{flushright} }

\medskip

\title{The Importance of Quantum Pressure of Fuzzy Dark Matter on Lyman-Alpha Forest}

\author{Jiajun Zhang$^{a,b}$\footnote{liamzhang@sjtu.edu.cn}}
\author{Jui-Lin Kuo$^c$\footnote{juilinkuo@gapp.nthu.edu.tw}}
\author{Hantao Liu$^{b}$\footnote{htliu@phy.cuhk.edu.hk}}
\author{\\Yue-Lin Sming Tsai$^d$\footnote{smingtsai@gate.sinica.edu.tw}}
\author{Kingman Cheung$^{c,d,e}$\footnote{cheung@phys.nthu.edu.tw}}
\author{Ming-Chung Chu$^{b}$\footnote{mcchu@phy.cuhk.edu.hk}}
\affiliation{$^a$ Department of Astronomy, School of Physics and Astronomy, Shanghai Jiao Tong University, Shanghai, China,}
\affiliation{$^b$ Department of Physics, The Chinese University of Hong Kong, Hong Kong, China,}
\affiliation{$^c$ Department of Physics, National Tsing Hua University, Hsinchu, Taiwan,}
\affiliation{$^d$ Physics Division, National Center for Theoretical Sciences, Hsinchu, Taiwan,}
\affiliation{$^e$ Division of Quantum Phases and Devices, School of Physics, Konkuk University,
Seoul 143-701, Republic of Korea}

\begin{abstract}
With recent Lyman-alpha forest data from BOSS and XQ-100,  some studies suggested that the lower mass limit on the fuzzy dark matter (FDM) particles is lifted up to $10^{-21}\ev$. However, such a limit was obtained by $\Lambda$CDM simulations with the FDM initial condition and the quantum pressure (QP) was not taken into account which could have generated non-trivial effects in large scales structures.
We investigate the QP effects in cosmological simulations systematically, and find that the QP leads to further suppression of the matter power spectrum at small scales, as well as the halo mass function in the low mass end. We estimate the suppressing effect of QP in the 1D flux power spectrum of Lyman-alpha forest and compare it with data from BOSS and XQ-100. The rough uncertainties of thermal gas properties in the flux power spectrum model calculation were discussed. We conclude that more systematic studies, especially with QP taken into account, are necessary to constrain FDM particle mass using Lyman-alpha forest.


\end{abstract}
\date{\today}


\maketitle

\section{Introduction}

Dark matter is one of the intriguing mysteries of modern
cosmology. Currently, the leading paradigm of dark matter is the cold dark
matter (CDM), supported by the majority of the observations like the
mass-to-light ratio of clusters of galaxies~\citep{MLratio}, the
rotation curves of galaxies~\citep{einasto1974dynamical}, the Bullet
Cluster~\citep{Bullet}, the cosmic microwave background
(CMB)~\citep{Planck} and the large scale structure of the
universe~\citep{MatterPowerSpectrum}. Despite its success on large
scales, the CDM paradigm faces three problems on small scales, dubbed as the
``small scales crisis''~\citep{SmallScaleCrisis}: (i) the missing
satellite problem, (ii) the cusp-core problem, and (iii) the too-big-to-fail 
problem. The essence of these problems is that CDM predicts an
excess amount of dark matter on small scales, and hence the key to address
them is to smooth out the small-scale structures by astrophysical
processes~\citep{AstroSolSSC}, or invoke alternative dark matter
models like warm dark matter (WDM)~\citep{WDM}, self-interacting dark
matter~\citep{SIDM} and fuzzy dark matter (FDM)~\citep{FDM}.

The FDM paradigm, in which the dark matter is made of ultra-light bosons
in Bose-Einstein condensate state, is an ideal alternative of CDM, 
since it suppresses small-scale structures 
while inherits the success of CDM on large scales
\citep{ULAReview, du2016substructure,Mocz:2017wlg}.
For the detailed history and implementation of FDM, one can see Ref.~\citep{Lee:2017qve} and the reference therein.
The suppression effect, arising from the effective ``quantum pressure'' (QP) 
of FDM, is directly connected to the mass of the ultra-light axion. 
The predictions of FDM with mass $\sim 10^{-22}\ev$ are 
consistent with observations of 
the CMB and large scales structure~\citep{FDMCMBLSS}, 
high-$z$ galaxies and CMB optical depth~\citep{FDMHMF}, and the
density profiles of dwarf spheroidal galaxies~\citep{FDMSim}.
However, recent results claimed that FDM with mass 
below $10^{-21}\ev$ 
has already been  ruled out at $95\%$ confidence level 
by comparing the results of hydrodynamic simulations to the 
Lyman-alpha forest data~\citep{irvsivc2017first, armengaud2017constraining, 2017arXiv170800015K}.


Lyman-alpha forest, a series of absorption lines in the Lyman $\alpha$
emission spectrum from distant galaxies and quasars by neutral hydrogen (HI) gas
at different redshift, provides the information about the the spatial
distribution of HI gas at high redshift. In the leading theory of
Lyman-alpha forest --- the gravitational confinement model --- HI
clumps are confined by the gravity provided by dark matter
halos~\citep{GravConf1, GravConf2}.  Thus, the flux power spectrum of
Lyman-alpha forest is a biased representation of the underling DM
density field power
spectrum~\citep{weinberg2003lyman,arinyo2015non,Viel:2004bf}.
Recently, two collaborations, the Baryon Oscillation Spectroscopic Survey
(BOSS)~\citep{SDSS2013lyman} and XQ-100~\citep{2016A&A...594A..91L}, have announced their analyses between redshift
$z=2-5$ for the flux power spectrum, thus providing tools to constraint
dark matter models on an unprecedented high level of precision.

To robustly exclude such a mass range, two tasks
have to be carried out before the experimental data analysis. The first
is numerical simulations of the FDM system on
large scales, which have been performed in a number of approaches: 
directly solving the
Schr\"odinger-Poission system~\citep{FDMSim}, Smoothed Particle
Hydrodynamic (SPH) simulation~\citep{mocz2015numerical}, and N-body
simulation with Particle-Mesh (PM) method~\citep{veltmaat2016cosmological}. 
However, all these simulations were
restricted in simulation scale and suffered from the singularity problem at
zero-density points in the calculation of QP. In this paper, we adopt
an independent simulation scheme developed in
Ref.~\citep{zhang2016ultra}, which provides DM-only simulations with
SPH and particle-particle (PP)
interactions to account for QP, is feasible for implementation in
cosmological scale simulations, and avoids the singularity problem.

The second is the uncertainties in hydrodynamic
simulations.  In the N-body simulation, the uncertainty in the matter power spectrum is
$\mc{O}(10\%)$ originating from the use of different initial
conditions, the adoption of different orders of Lagrangian perturbation
theory~\citep{l2014effects}, finite-box
effect~\citep{bagla2006effects} and the usage of different N-body
simulation
codes~\citep{heitmann2010coyote,kim2013agora,sembolini2016nifty}. These
uncertainties can be precisely estimated and controlled because 
their properties  are well studied.  In contrast, the uncertainties from hydrodynamic simulations can be of
orders of magnitude~\citep{o2005comparing,vazza2011comparison,kim2013agora,sembolini2016nifty}.
Different treatments introduced in hydrodynamic simulations in handling
the gaseous part induce huge discrepancy in gas density, gas
temperature and galaxy structure.  Moreover, different hydrodynamic
simulation codes contribute uncertainties to the 1D flux power spectrum
at the level of $5\%$.  As stated in Ref.~\citep{Hui:2016ltb}, several
additional astrophysical processes may also alter the 1D flux power
spectrum, such as patchy reionization and smoothing.

We estimate the constraint on the mass range of
FDM from Lyman-alpha forest data with DM-only cosmological
simulations in four different settings: (1) standard CDM simulation,
(2) simulation using FDM-modified initial condition  and CDM dynamics, (3) simulation
using FDM-modified initial condition and dynamics with FDM mass equals
to $2.5\times10^{-22}\ev$, (4) $2.5\times10^{-23}\ev$.  By
utilizing the 1D flux power spectrum of the CDM hydrodynamic simulation in Ref.~\citep{irvsivc2017first}, 
one can obtain the 1D flux
power spectrum for the corresponding condition using the 3D power
spectrum ratio from our simulations.

Our results are shown in the aspects of density field, halo mass
function, 3D power spectrum and 1D flux power spectrum with
Lyman-alpha forest data. We show that the effect of QP in structure formation
is non-trivial and introduce important suppression in the growth of structures.
Inspired by the linear theory summarized in Ref.~\citep{mo2010galaxy}, 
we have calculated the uncertainty range of the 1D flux power spectrum 
based on the unknown hydrogen gas temperature properties. 
The suppression of the 1D flux power spectrum 
from higher gas temperature behaves like the suppression caused by 
FDM. 


This paper is arranged as follows. 
In Sec.~\ref{sec:method}, we discuss the methodology of the simulation, simulation set-up and initial conditions. 
We also review the linear theory of 1D flux power spectrum calculation.
In Sec.~\ref{Sec:result}, we present our results in the aspects of density field, 
halo mass function, 3D power spectrum and 1D flux power spectrum.
We compare our 1D flux power spectrum to the Lyman-alpha forest data.
In Sec.~\ref{sec:discussion}, we discuss the gas temperature uncertainty.
Finally, we summarize our outcomes in Sec.~\ref{Sec:conclusion}.

\section{Methodology}\label{sec:method}

\subsection{Quantum Pressure as particle-particle interaction}

Standard N-body simulations have problems in calculating the QP
because of the discretization of the density field by the delta
function. From the expression of QP, obviously, what we need in the
calculation is a smooth density field. There are already some reliable
smoothing particle methods used in the N-body simulation of the FDM system
with limited box sizes ($<10h^{-1}\Mpc$), e.g.,
Ref.~\citep{mocz2015numerical, veltmaat2016cosmological}. For the sake
of Lyman $\alpha$ forest, it is important to have a cosmological scale
simulation so that the study of structure formation and matter power
spectrum on large scales is possible.

We proposed a novel N-body simulation scheme for FDM
in a previous work~\citep{zhang2016ultra}, in which the delta function is
replaced by a smooth Gaussian kernel function to solve the singularity
problem and avoid the simulation crash. This scheme can be used to perform simulations at scales no smaller than $50h^{-1}\Mpc$ as we
shall prove later and give a coarse-grained description on
 small scales. However,
the original scheme requires some modifications for cosmological
simulations. In this section, we will demonstrate how to embed QP in
the cosmological simulations.

{
For cosmological simulations, one has to consider the
transformation from physical to comoving coordinates. 
Under the transformation described in Appendix~\ref{app:comoving}, 
there is an additional pre-factor $a^{-2}$ for the original QP defined in Ref.~\citep{zhang2016ultra}, where $a$ is the cosmological scale factor. 
}
The QP for a cosmological simulation becomes 
\begin{equation}
	\label{eq:QPcomoving}
	Q=-\dfrac{\hbar^2}{2m_\chi^2 a^2}\dfrac{\bm{\nabla}^2\sqrt{\rho}}{\sqrt{\rho}},
\end{equation}
, where $\hbar$, $m_{\chi}$, and $\rho$ are the reduced Planck constant, FDM particle mass and the mass density of FDM, respectively.
The corresponding acceleration can be written as
\begin{equation}
	\label{eq:accelcomoving}
	\ddot{\bm{r}}=\dfrac{4M\hbar^2}{M_0 m_\chi^2 \lambda^4 a^2}\sum_j\mathcal{B}_j\exp
	\left[-\dfrac{2|\bm{r}-\bm{r_j}|^2}{\lambda^2}\right]
	\left(1-\dfrac{2|\bm{r}-\bm{r_j}|^2}{\lambda^2}\right)(\bm{r_j}-\bm{r}),
\end{equation}
where $M$, $M_0$, $\lambda$, and $\mathcal{B}_j$ are the mass of the
simulation particle, a normalization factor accounting for the volume
$\Delta V_{j}$ occupied by simulation particles, the de Broglie
wavelength of FDM  particles, and the correction factor for high-density regions,
respectively. For more detailed explanations of $\mathcal{B}_j$, please refer to Appendix~A of Ref.~\citep{zhang2016ultra}. 



\begin{figure}
\centering
\includegraphics[scale=0.5]{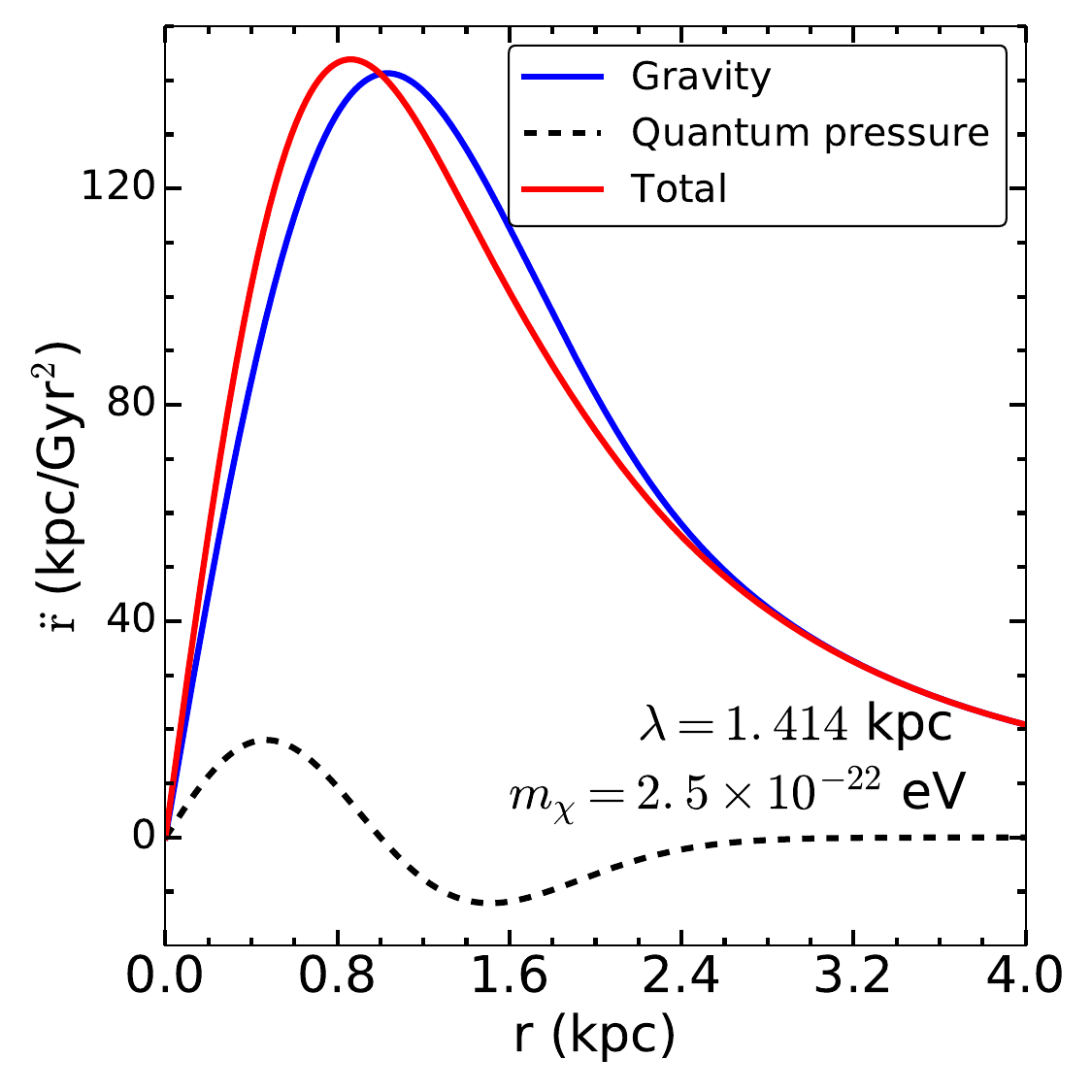}
\caption{The acceleration from QP (black dashed line), gravity (blue solid line), and their sum (red solid line)  vs.  the distance between two particles. Notice that QP is at least 1 order of magnitude smaller than the gravity in the simulation.}
\label{Fig:quantumpressure}
\end{figure}

To demonstrate the effect of QP, we consider a two-particle system
separated by a distance of order $\mathcal{O}(\mathrm{kpc})$, and the
acceleration caused by QP will be
$\mathcal{O}(\hbar^2\,m_\chi^{-2}\,\lambda^{-3})\sim\mathcal{O}(10^{-10}\,\text{m}\,\text{s}^{-2})$.
In Fig.~\ref{Fig:quantumpressure}, the QP effect is presented in the
plane of ($r$, $\ddot{r}$). The acceleration from QP, gravity, and their
sum are shown by the black dashed line, blue solid line, and red solid
line, respectively.  Apparently, the effect of QP is attractive if the
separation between the two particles is shorter than
$\lambda/\sqrt{2}$, otherwise repulsive.  To understand this
phenomenon, we refer back to the definition of QP in Eq.~\eqref{eq:QPcomoving}:
$Q$ is proportional to the curvature of the density, which can be
negative, positive, or zero, physically corresponding to negative,
positive, and zero forces, respectively.  The reason of the difference
between Fig.~\ref{Fig:quantumpressure} here and the Fig.~1
of~\citep{zhang2016ultra} is that the mass of the simulation particle
is changed from $10^6M_\odot$ to $\sim 10^{8}M_\odot$.  The effect of
gravity is directly related to the mass of the simulation particles
while the QP is not affected due to the fact that the normalization factor
$M_0$ in Eq.~\eqref{eq:accelcomoving} is chosen to be the same as the mass of
the simulation particles. 
In other words,
the strength of QP is a quantity independent of the simulation particle mass 
$M$.  Therefore, the acceleration
of gravity is considerably larger in a cosmological simulation due to
the larger mass of the simulation particles. Thus, the
structure formation in large scales is still dominated by gravity. However,
the QP plays a non-trivial role in the structure formation, particularly in 
highly non-linear regions such as dark matter halos.

\subsection{Simulation settings}


We use the code \texttt{Gadget2}~\citep{gadget2}, which is a
\texttt{TreePM} hybrid N-body code, to perform our simulations. To
describe QP, we discretize the interaction term and modify
\texttt{Gadget2}\footnote{The code is called \texttt{Axion-Gadget}, publicly available at \url{https://github.com/liambx/Axion-Gadget}} to calculate QP in the same way as
in Ref.~\citep{zhang2016ultra}. (The PM method is helpful for
cosmological simulations with periodic boundary conditions.) Because QP
behaves like a short-range force, we adopt the original PM code to
compute the long-range force and modify the Tree code which takes care
of the short-range force to include the calculation of QP. In addition, it
is not necessary to set softening length for QP since QP is finite in
the $(\bm{r_j}-\bm{r})\simeq 0$ region.

We start our simulation from the redshift $z=99$.
The related cosmological parameters are DM energy density $\Omega_{m}=0.3$, 
cosmological constant $\Omega_{\Lambda}=0.7$, 
baryon energy density $\Omega_{b}=0.04$, 
dimensionless Hubble parameter $h=0.7$, 
scalar spectral index $n_s=0.96$,
and the power spectrum normalization factor $\sigma_{8}=0.8$.


\begin{figure}
\centering
\includegraphics[scale=0.5]{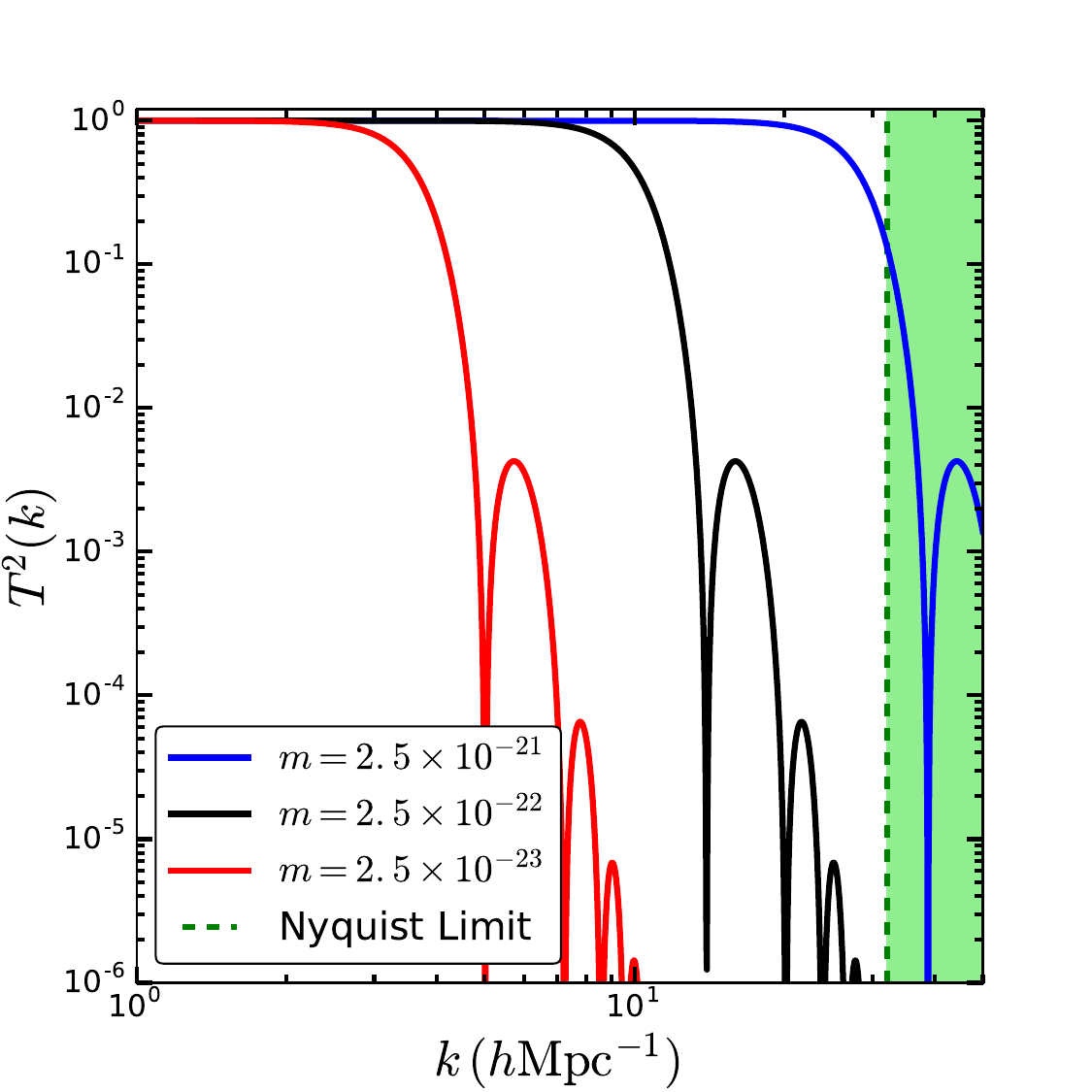}
\caption{Initial condition transfer function  squared  vs. the wavenumber. The blue, black, and red solid lines correspond to FDM masses of $2.5\times10^{-21}\ev$, $2.5\times10^{-22}\ev$, $2.5\times10^{-23}\ev$, respectively. The green dashed line is the Nyquist Limit which is the resolution limit of our simulation, and so the light green region is not trustworthy.
The suppression of the power spectrum in small scales depends on the mass of the FDM particles~\citep{Hu:2000ke}: the smaller the FDM particle mass, the larger the suppression scale will be.}
\label{Fig:transfer}
\end{figure}

We generated the initial CDM power spectrum following Ref.~\citep{eisenstein1998baryonic}.   
The suppression of the FDM power spectrum relative to the 
CDM one on small scales can be characterized by a transfer 
function $\mathcal{T}(k,z)$ \citep{Hu:2000ke}
\begin{equation}
	\label{eq:suppresion}
	P_{F}(k,z)=\left[\frac{\mathcal{T}_{\texttt{FDM}}(k,z)}{\mathcal{T}_{\texttt{CDM}}(k,z)}\right]^{2}P_{C}(k,z)
	=\mathcal{T}^{2}(k,z)P_{C}(k,z),
\end{equation}
where $k$ is the wavenumber and $P_{F}(k,z)$ and $P_{C}(k,z)$ are the 
three-dimensional power spectra of FDM and CDM, respectively.
One has to bear in mind that the physical difference between 
$\mathcal{T}_{\texttt{CDM}}(k,z)$ and $\mathcal{T}_{\texttt{FDM}}(k,z)$
\footnote{The transfer functions $\mathcal{T}_{\texttt{FDM}}(k)$ and
  $\mathcal{T}_{\texttt{CDM}}(k)$ are defined in Eq.~(1) of
  Ref.~\citep{eisenstein1998baryonic}, which transfer the power
  spectrum at large scales ($k=0$) to the power spectrum at other scales
  ($k\neq0$) for the corresponding model.
Restricted to linear theory, the current matter power spectrum $P(k)$
would be proportional to the product of the primordial power spectrum
$P_{prim}(k)$ given by inflation theory and $\mathcal{T}^2(k)$ embodying the
growth of structure, $P(k)\propto \mathcal{T}^2(k)P_{prim}(k)$. At the
super-horizon scale, cosmological perturbation theory shows that the
structure does not grow, and hence the transfer function is set with
boundary condition $\mathcal{T}(0)=1$.
}
is characterized by 
the transfer function $\mathcal{T}(k,z)$, which transforms
the power spectrum from CDM to FDM and so 
is deduced as the ratio of FDM transfer function $\mathcal{T}_{\texttt{FDM}}(k,z)$ to
CDM transfer function $\mathcal{T}_{\texttt{CDM}}(k,z)$.

Following the arguments in Ref.~\citep{Schive:2015kza}, we can well approximate $\mathcal{T}(k,z)$ by the redshift-independent expression~\citep{FDM} 
\begin{equation}\label{transfer}
\mathcal{T}(k)=\dfrac{\cos x^{3}}{1+x^8},~~{\rm where}~x=1.61\times\left(\frac{m_\chi}{10^{-22}\ev}\right)^{1/18}\times \frac{k}{k_{J}},
\end{equation}
The parameter $k_{J}=9(m_\chi/10^{-22}\ev)^{1/2}\mathrm{Mpc}^{-1}$ is 
the critical scale of Jeans wavenumber at matter-radiation equality.  In Fig.~\ref{Fig:transfer}, we present the square of
the transfer function $\mathcal{T}^{2}(k)$ with three different FDM masses to
demonstrate the suppression of FDM power spectrum in small scales
relative to CDM power spectrum. The blue, black, and red solid lines
correspond to the masses of the FDM being 
$2.5\times10^{-21}\ev$, $2.5\times10^{-22}\ev$,
$2.5\times10^{-23}\ev$, respectively.  The vertical green dashed line
represents the Nyquist limit which is the resolution limit of our
simulations and the corresponding wavenumber $k_{Ny}$ is computed to
be
\begin{equation}
k_{Ny}=\pi\left(\frac{N_0}{V_0}\right)^{1/3},\nonumber
\end{equation}
where $V_0$ is the volume of the simulation box, and $N_0$ is the total
number of simulation particles, which are $\left(50 h^{-1}\mathrm{Mpc}\right)^3$ and $512^3$ respectively in our simulations.
The origin of this limit is due to the fact that each simulation
particle has an approximate average volume $(V_0/N_0)^{1/3}$ and we cannot 
know what happens inside the simulation particle.  In other words,
the result is not reliable for $k>k_{Ny}$.  One can see that there is
a sharp break of FDM power spectrum at $k\sim k_{J}$ and severe
oscillations occur with suppression for $k>k_{J}$ for all three different
FDM masses. For a smaller mass, the suppression is tremendous compared
to a larger mass which behaves almost the same as the CDM case at high
$k$.

We modify the code \texttt{2LPTic}~\citep{2lptic}
for generating the initial conditions of the CDM power spectrum for 
cosmological simulations to incorporate the transfer function so as to generate initial conditions with the FDM power spectrum.

\subsection{1D Flux Power Spectrum}
\label{subsec:DataAnalysis}
The calculation of the  1D flux power spectrum in the linear regime is well summarized in Ref.~\citep{mo2010galaxy}, 
\bea
	P_b&=&\f{P_{DM}}{\left(1+k^2/k_J^2\right)^2},\label{eq:PDMtoPb}\\
	P_F(k_z,z)&=&\int_{k_z}^{\infty}\dfrac{k\dd k}{2\pi}P_{b}(k,z)W(k,k_z).\label{eq:3Dto1D'}
\eea
First, the linear dark matter power spectrum $P_{DM}$ was calculated by \texttt{CAMB}~\citep{lewis2011camb}, with the same parameters used in the simulations. 
Then we calculate the baryon power spectrum from Eq.~\eqref{eq:PDMtoPb}, in which $k_J$ is the Jeans wavenumber related to the Jeans length $\lambda_J$ by
\be
	k_J=\f{2\pi}{\lambda_J}.
\ee
Finally, we integrate Eq.~\eqref{eq:3Dto1D'} to get the one-dimensional flux power spectrum, in which we need the bias function
\be
	\label{eq:LinearBiasFunc}
	W=A\exp\left(-\f{k_z^2b_0^2}{2H^2}\right)\left[1+\f{\Omega_m^{0.6}}{2+0.7\left(1-\gamma\right)}\f{k_z^2}{k^2}-\f{\gamma-1}{4\left[2+0.7\left(1-\gamma\right)\right]}\f{k_z^2b_0^2}{H^2}\right]^2.
\ee
Here 
\be
	b_0(z)=\sqrt{\f{2k_BT_0(z)}{m_H}} \nonumber
\ee
is a parameter related to the velocity dispersion of the HI gas at redshift $z$, and $T_0(z)$ is the average temperature of HI gas at redshift $z$. $\gamma(z)$ is the polytropic index in the equation of state of HI gas
\footnote{
For an ideal polytropic gas, the temperature at redshift $z$ and position $\bm{x}$ is 
\ben
	T(z,\bm{x})=T_0(z)\left(1+\delta_b(\bm{x})\right)^{\gamma(z)-1}.
\een
}.
$A$ is a normalizing factor, 
whose effect is the same as the parameter $\sigma_8$ in $P_{DM}$.

For further calculation, we use a power-law parametrization of $T(z)$ and $\gamma(z)$
\be
	\label{eq:PowerLawParameterization}
	T_0(z)=T_0^A\left[\f{1+z}{5.5}\right]^{T_0^S},\quad\gamma(z)=\gamma^A\left[\f{1+z}{5.5}\right]^{\gamma^S},
\ee
where the 1$\sigma$ uncertainty ranges are
\be
	\label{eq:ThermalParameter}
	T_0^A=9.2_{-0.1}^{+1.2}10^3\,\text{K},
	\quad T_0^S=-2.5_{-0.5}^{+0.45},
	\quad\gamma^A=1.64_{-0.26}^{+0.01},
	\quad \gamma^S=-0.15_{-0.61}^{+1.25}.
\ee
These are copied from the Table.~II in Ref.~\cite{viel2013warm}. As for $\sigma_8$ and $\lambda_J$, we treat them as redshift independent constants with the 1$\sigma$ uncertainty ranges as in Ref.~\cite{Planck, rorai2017measurement}
\be
	\sigma_8\in\left[0.78, 0.88\right],\quad\lambda_J=100\pm80\kpc.
\ee

The formulae discussed above are only applicable to linear perturbation, and the insertion of non-linear matter power spectrum in Eq.~\eqref{eq:3Dto1D'} is problematic. However, for the sake of simplicity, the 1D flux power spectrum is assumed to be related to the DM density power spectrum with the following equation,
\begin{equation}\label{eq:3Dto1D}
P_F(k_z,z)=\int_{k_z}^{\infty}\dfrac{k\dd k}{2\pi}P_{DM}(k,z)W(k),
\end{equation} 
which is inspired by the linear theory. Here $W$ represents the bias introduced by the thermal properties of intergalatic medium (IGM), but the distortions in the direction of line-of-sight is neglected.
Given the form of  Eq.~\eqref{eq:3Dto1D}, the 1D flux power spectrum is the 3D matter power spectrum convolved with the bias $W$ along the line of sight.
On one hand, we assume that the dynamics of dark matter does not affect the gaseous component, and the bias $W$ is the same in all four different simulations. On the other hand, we reweigh the matter power spectrum $P_{DM}$ with the 3D matter power spectrum ratio of the simulations \textbf{FDM}/\textbf{FIC}/\textbf{F23} to the simulation \textbf{CDM}. That is to say, we first take the derivative of the 1D flux power spectrum $P_F$, from the mock 1D flux power spectrum in hydrodynamic simulation shown in Ref.~\citep{Irsic:2017sop}, with respect to $k_z$
\be
	\label{eq:DerivativePowerSp}
 	-\f{2\pi}{k_z}\f{\dd P_F}{\dd k_z}=P_{DM}W,
\ee
and multiply the right hand side (RHS) of this equation by the ratio $P_{i}/P_{\text{CDM}}$, where $i=$ \textbf{FDM}, \textbf{FIC} and \textbf{F23}. Then we integrate this expression to get the modified 1D flux power spectrum
\be
	P_{F,i}(k_z,z)=\int_{k_z}^\infty\f{k\dd k}{2\pi}\f{P_{i}(k,z)}{P_{\text{CDM}}(k,z)}P_{DM}(k,z)W(k).
\ee
By comparing the modified $P_{F,i}$ and the original $P_F$, we can tell the effect of FDM model on Lyman alpha forest.

\section{Numerical result}\label{Sec:result}

\begin{table}[h]
\begin{center}
\begin{tabular}{p{2.5cm}<{\centering}|p{3.5cm}<{\centering}|p{3.5cm}<{\centering}|p{3.5cm}<{\centering}}
\hline
\hline
Abbreviations & Initial Conditions & Dynamics & FDM mass $m_\chi$\\
\hline
\textbf{CDM} & CDM-Standard & CDM-Standard & ---\\ 
\textbf{FIC} & FDM-modified & CDM-Standard &$2.5\times 10^{-22}\ev$\\ 
\textbf{FDM} & FDM-modified & FDM-modified & $2.5\times 10^{-22}\ev$\\ 
\textbf{F23} & FDM-modified & FDM-modified & $2.5\times 10^{-23}\ev$\\
\hline
\hline
\end{tabular}
\caption{The abbreviations and details of simulations we have performed.}
\label{Tab:simulation}
\end{center}
\end{table}

We have performed four different kinds of simulations for 
comparison as listed in Table.~\ref{Tab:simulation}.
To avoid words cluttering in the following presentation, 
we use abbreviations for these four simulations.
In this section, we first present the density field of the simulation 
\textbf{FDM} 
as well as  
the density field difference between the simulations 
\textbf{FDM} and \textbf{FIC}.
We then compare the 3D power spectra and 1D flux power spectra of all four simulations in order to investigate the non-linear effect of the QP. 
After showing the 1D flux power spectrum, 
a numerical chi-square test of Lyman-alpha forest data from BOSS and XQ-100 
is performed.
Finally, we discuss the constraint on the FDM mass.

\subsection{Density Field and Halo Mass Function}

\begin{figure}
\centering
\includegraphics[width=0.45\textwidth]{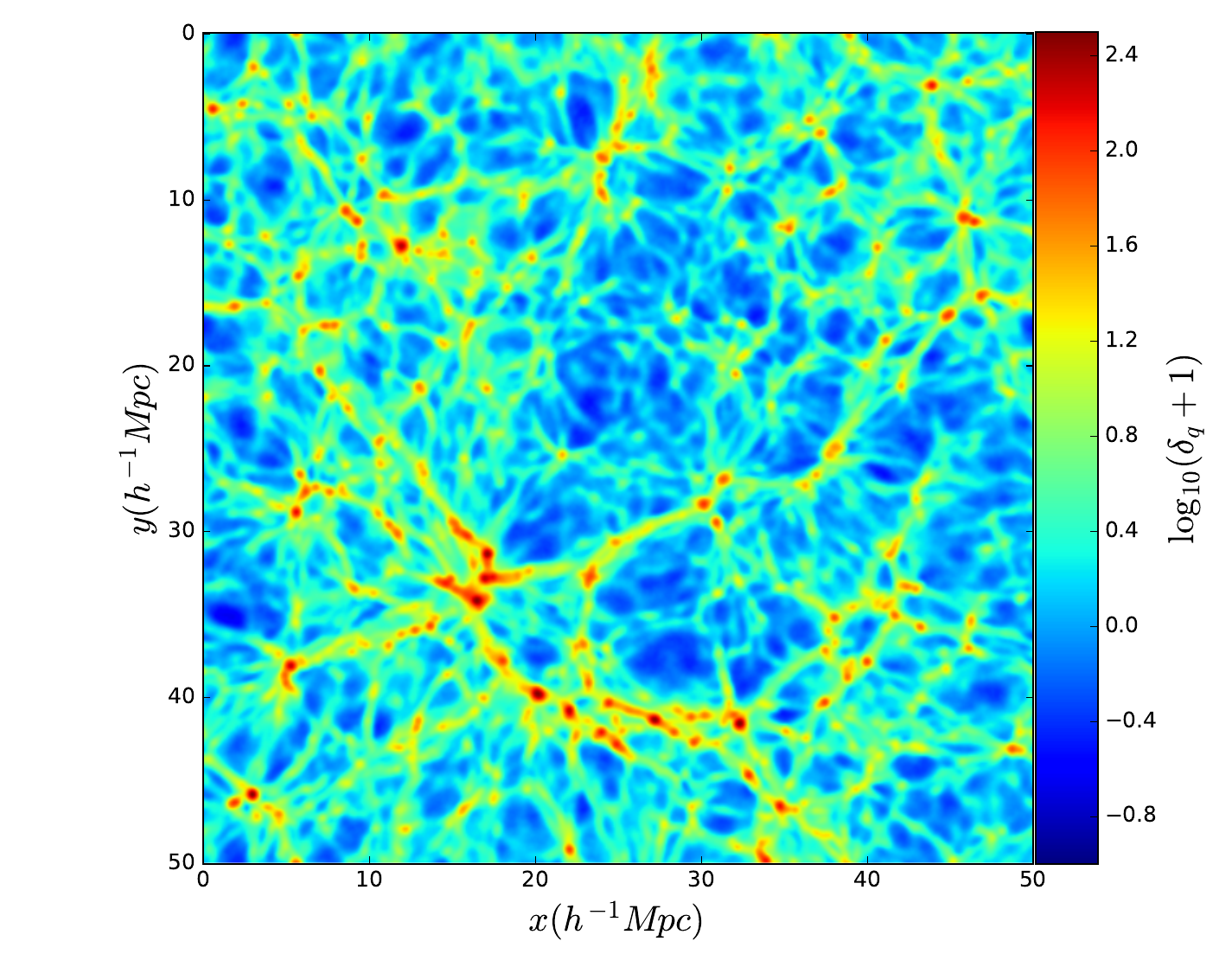}
\includegraphics[width=0.45\textwidth]{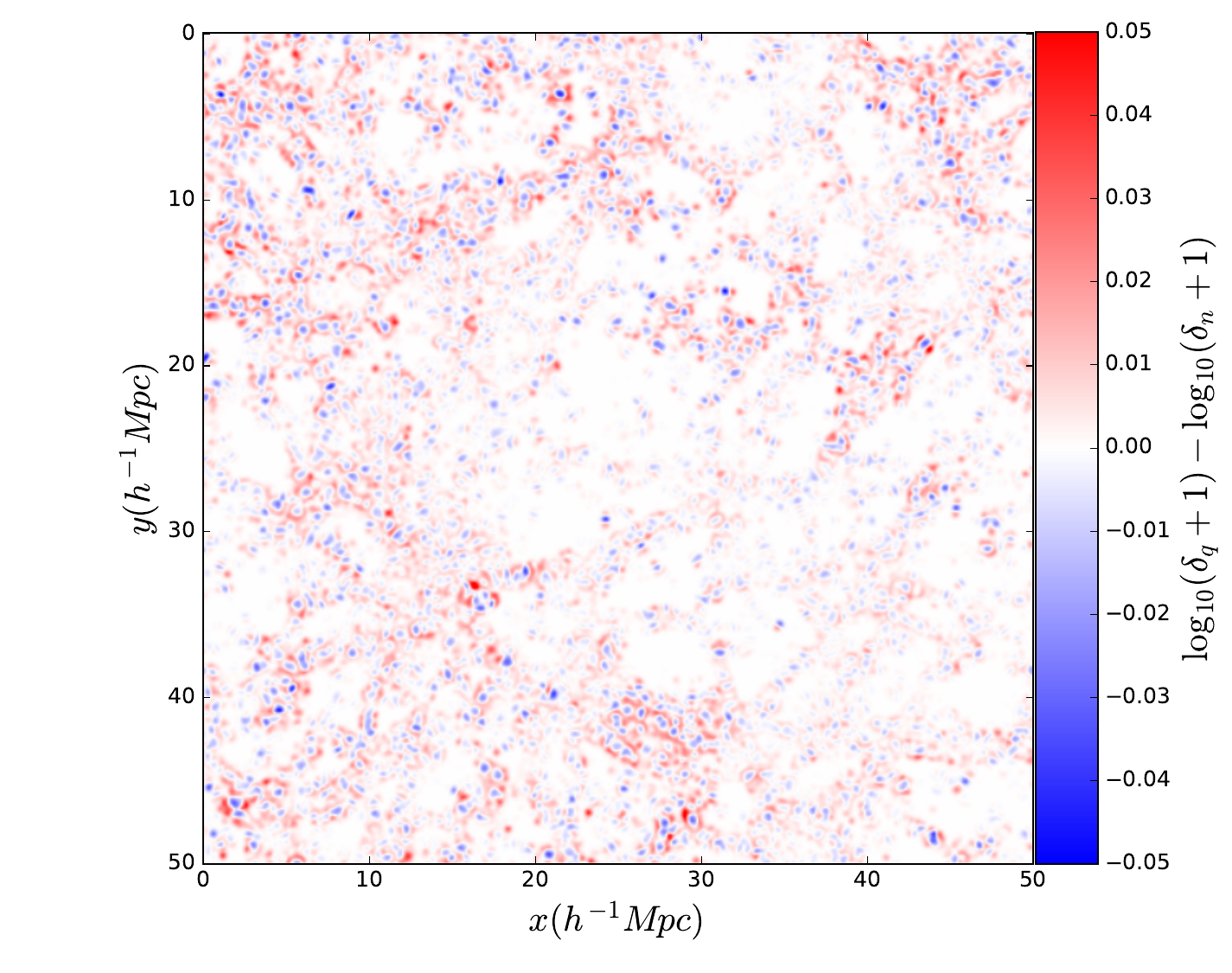}
\caption{Left panel: the density field of the simulation \textbf{FDM}. This figure is a slice from the simulation cube. The color scale represents the log scale of density contrast $\delta_{q}$. Right panel:  the difference of density field between the simulation \textbf{FDM} and the simulation \textbf{FIC}. The color scale represents the log scale of the density difference between the mentioned two simulations. $\delta_{n}$ is the density contrast of the simulation \textbf{FIC}.}
\label{Fig:dfield}
\end{figure}

In the left panel of Fig.~\ref{Fig:dfield}, we depict the density field 
taken from a slice of the simulation cube. 
The slice is $0.5\,h^{-1}\,\mathrm{Mpc}$ thick, and the density 
field is calculated from the 
particle distribution by the triangular shaped cloud (TSC) scheme 
but further smoothed by a Gaussian filter 
(variance $\sigma=0.15\,h^{-1}\,\mathrm{Mpc}$) for better illustration. 
With the color scale, one can clearly see that the 
large scales structures include voids, 
filaments and knots.
The density contrast $\delta_{q}$ is defined as
\begin{equation}
	\label{overdensity}
	\delta_{q}=\dfrac{\rho_{q}}{\bar{\rho_{q}}}-1,\nonumber
\end{equation}
where the subscript $q$ denotes that it is from the simulation \textbf{FDM}.
The color scale of Fig.~\ref{Fig:dfield} represents the logarithm of 
$\delta_{q}+1$, which is
the ratio of local density $\rho_{q}$ to the average density $\bar{\rho_{q}}$.
In the right panel of Fig.~\ref{Fig:dfield}, we compare the density field of the 
simulation \textbf{FDM} and the simulation \textbf{FIC}.
The color scale shows the difference between the logarithm of $\rho_{q}/\bar{\rho}_{q}$ and 
the logarithm of $\rho_{n}/\bar{\rho}_{n}$. 

We found that the difference between the density fields of the two simulations 
\textbf{FDM} and \textbf{FIC} is small, but the effect of QP can 
produce  granular structures 
close to the high density regions. 
Note that these granular structures are indeed due to QP,
but not the difference in the initial conditions by comparing the 
density fields of 
the simulations \textbf{FIC} and \textbf{CDM}.

\begin{figure}
\centering
\includegraphics[width=0.45\textwidth]{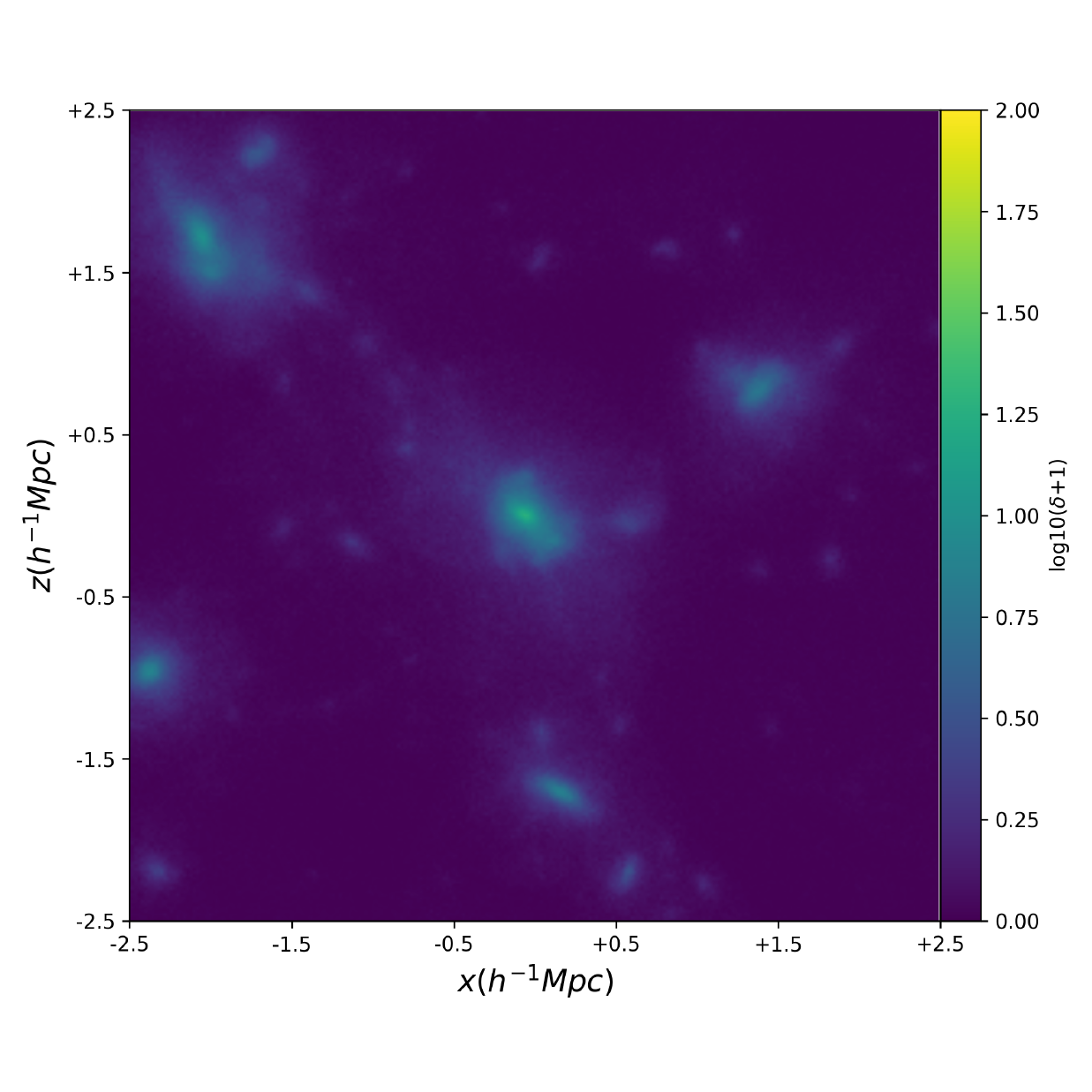}
\includegraphics[width=0.45\textwidth]{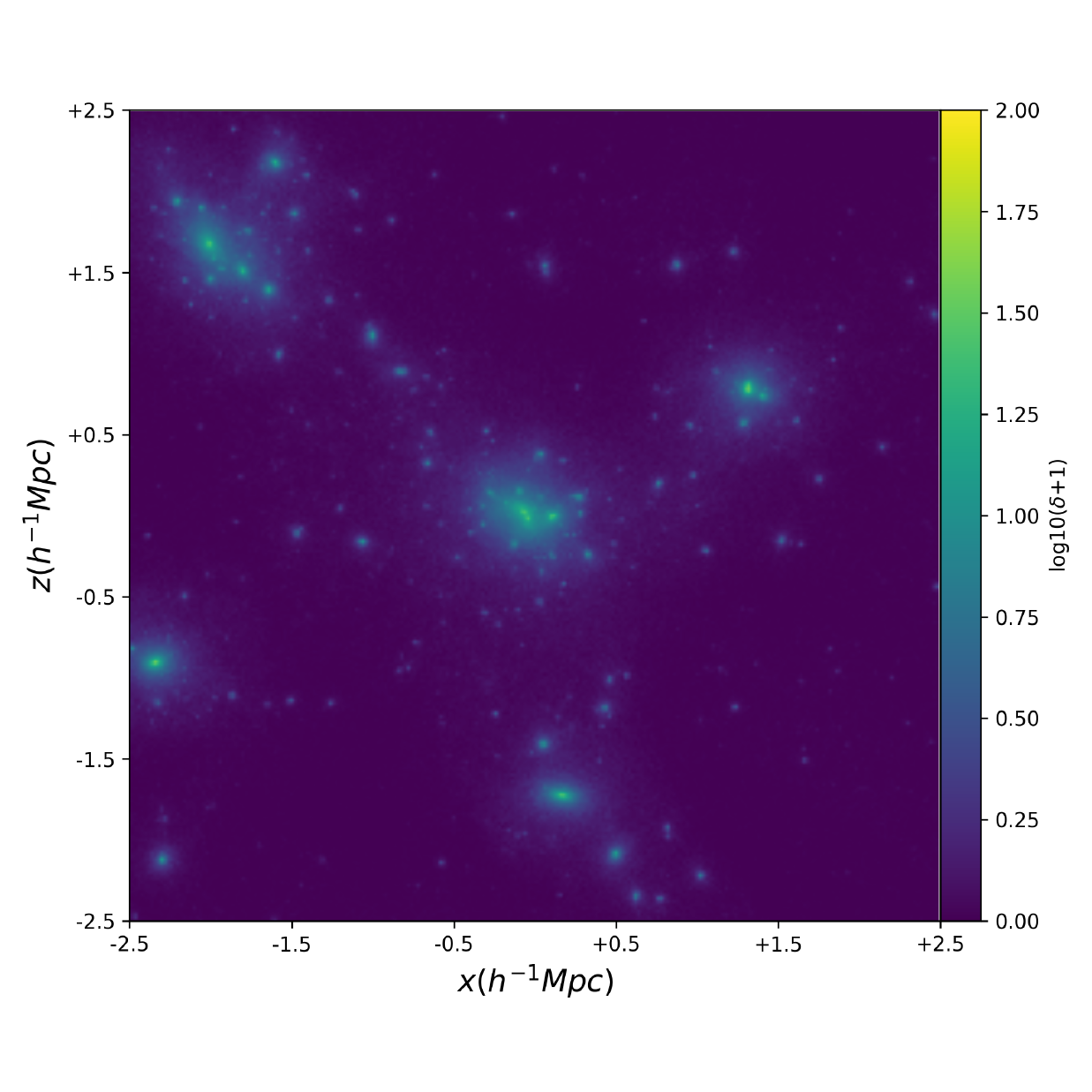}
\caption{{A $5h^{-1}$Mpc sub-box density field projected in y direction from the simulation \textbf{FDM} (on the left) and the simulation \textbf{FIC} (on the right). The color scale represents the log scale of density.
The substructure on the left panel (\textbf{FDM}) is suppressed under the influence of QP.}}
\label{Fig:cluster}
\end{figure}

{
To illustrate more details in small scale, we select 
a $5h^{-1}$Mpc sub-box projected in y direction
and zoom in for better resolution. 
In Fig.~\ref{Fig:cluster}, the left panel is from the simulation \textbf{FDM}
and the right panel is from the simulation \textbf{FIC}.
The density field is calculated from the particle distribution by the nearest grid point scheme and further smoothed by bilinear algorithm.
}
We note that the cluster in the simulation \textbf{FDM} looks 
much fuzzier than the one in the simulation \textbf{FIC}.
Qualitatively, the formation of low mass halos is further suppressed 
when taking QP into consideration.

\begin{figure}
\centering
\includegraphics[width=0.45\textwidth]{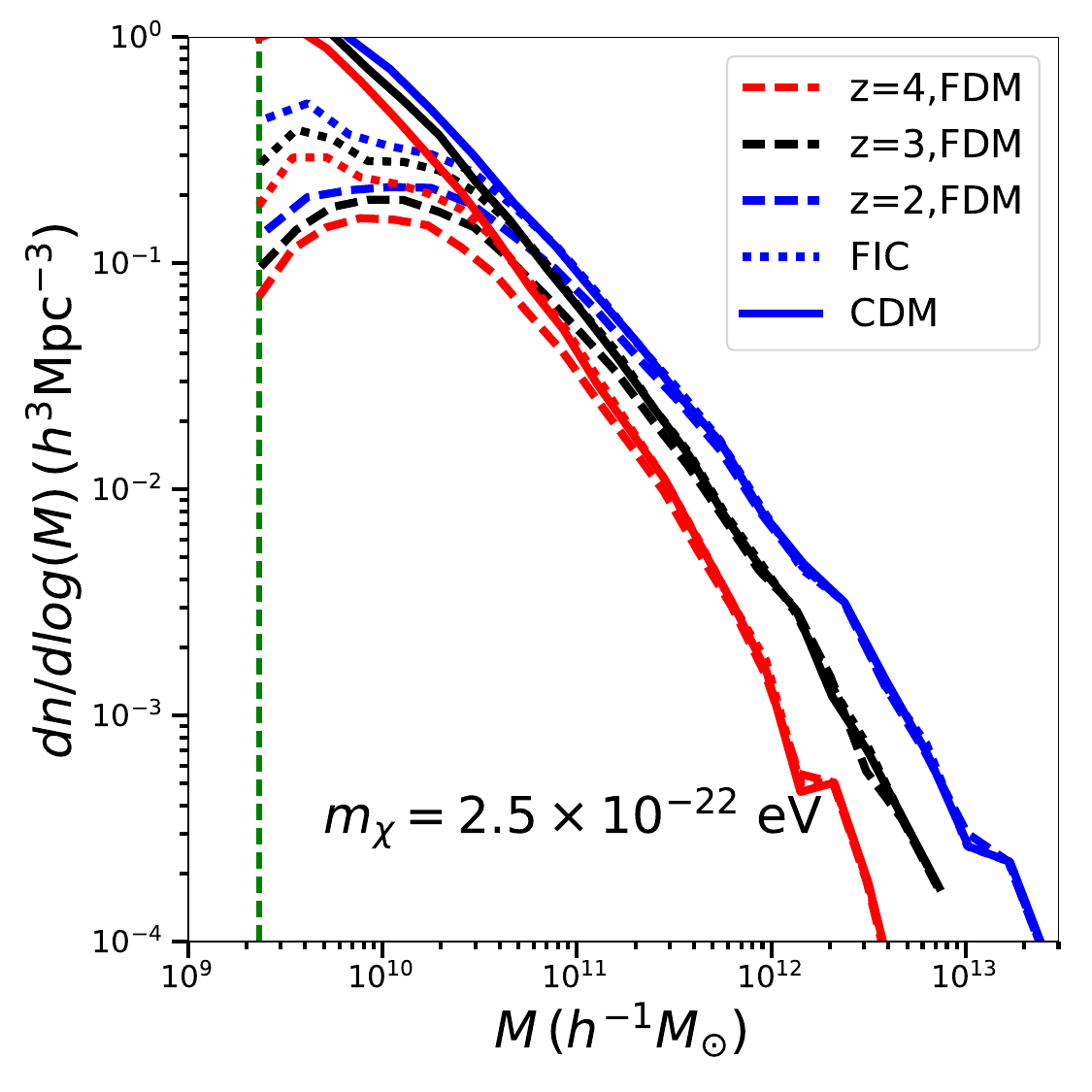}
\includegraphics[width=0.45\textwidth]{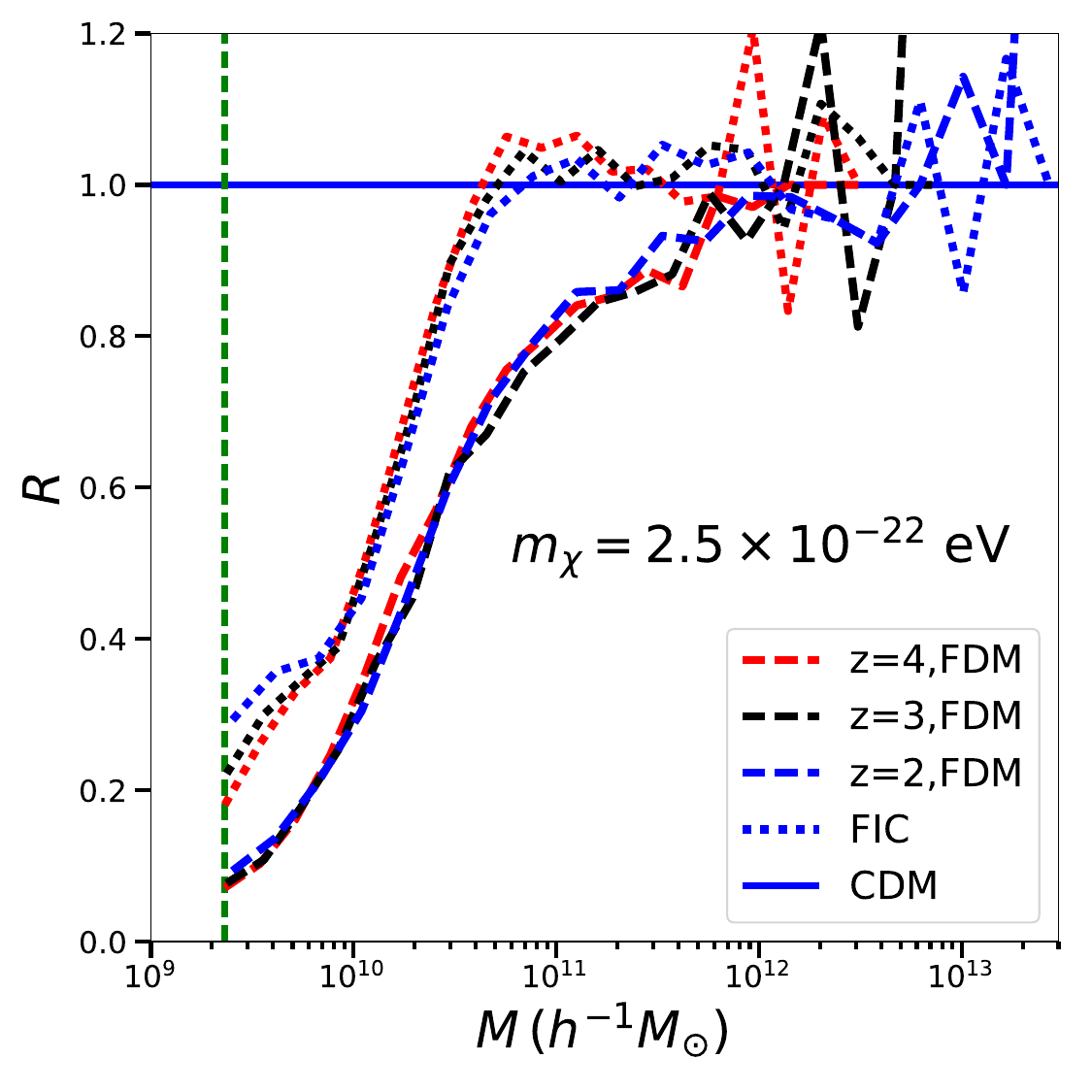}
\caption{Left panel: The halo mass function measured from the simulation \textbf{FDM} (dashed line), \textbf{CDM} (solid line) and \textbf{FIC} (dotted line) for redshift $z=4, 3, 2$ (red, black, blue). 
{
Right panel: The respective halo mass function ratios between the simulation \textbf{FDM} (dashed line), \textbf{FIC} (dotted line) and \textbf{CDM} for redshift $z=4, 3, 2$ (red, black, blue). 
Here the FDM mass is $m_\chi=2.5\times10^{-22}\ev$.
The abscissa axis is the mass of the halo and the ordinate axis is the number density per halo mass in log scale. The number density of low mass halos is further suppressed under the influence of QP by $\sim20\%$, if we compare the halo mass function ratio of \textbf{FDM} with that of \textbf{FIC}.} }
\label{Fig:hmf}
\end{figure}

%
%
%
{
We identified the halos using the package $\texttt{AHF}$~\cite{Knollmann:2009pb},
which can build up the hierarchical structure for the halos
and sub-halos in the snapshots of our simulations. The halos are identified if the
average density of the halo is over 200 times the critical density of the universe.
}
%
In Fig.~\ref{Fig:hmf}, 
the halo mass function is presented with the FDM mass
$m_\chi=2.5\times10^{-22}\ev$.  The colors represent different
redshifts, and the solid, dot-dashed and dashed lines represent the halo
mass function of the simulations \textbf{CDM}, \textbf{FIC} and
\textbf{FDM}, respectively.  The break of the halo mass functions at
$M=2.3\times10^9~h^{-1}\,{M_\odot}$ (the green dashed line) is due to our
limited resolution.  By definition, we cannot identify halos whose
mass is below $2.3\times10^{9}\,h\,{^{-1}M_\odot}$.  
From halo masses $5\times10^{11}\,h\,{^{-1}M_{\odot}}$ to
$2\times10^{13}\,h\,{^{-1}M_{\odot}}$, there is no recognizable
difference among the simulations \textbf{FDM}, \textbf{FIC} and
\textbf{CDM}. 
 However, for halo mass below
$5\times10^{11}\,h\,{^{-1}M_{\odot}}$, the difference becomes noticeable.

{
To quantitatively see the difference, one can refer to the right panel of Fig.~\ref{Fig:hmf} which manifestly demonstrates the suppression caused by the QP and the modified initial condition. We can see that QP (\textbf{FDM} simulation) introduces $20\%$ more suppression on the number density of $M<2\times10^{11} h^{-1}M_{\odot}$ halos than that of \textbf{FIC} simulation, with modified initial condition only. There is no identifiable effect in the simulations \textbf{FDM} and \textbf{FIC} for the halo mass function with $M>5\times10^{11} h^{-1}M_{\odot}$.
}

It is clear that the difference in modified initial conditions and QP start
to have significant influence on the formation of halos. 
Furthermore, it is worth mentioning that by using the code established in
Ref.~\citep{zhang2016ultra} we are able to explore halo mass
smaller than $2\times10^{13}\,h\,{^{-1}M_{\odot}}$ in
cosmological simulations.

\subsection{Impacts on Lyman-Alpha Forest}
\label{subsec:ImpactLyalpha}

Here we investigate the difference in the 1D flux power spectrum among our four simulations with the method mentioned in Sec.\ref{subsec:DataAnalysis}, 
which enables us to quantitatively study the effect of the modified initial condition and dynamics introduced by FDM. Furthermore, our result of 1D flux power spectrum will be used to compare with the BOSS/XQ-100 data.

\begin{figure}
\includegraphics[width=0.45\textwidth]{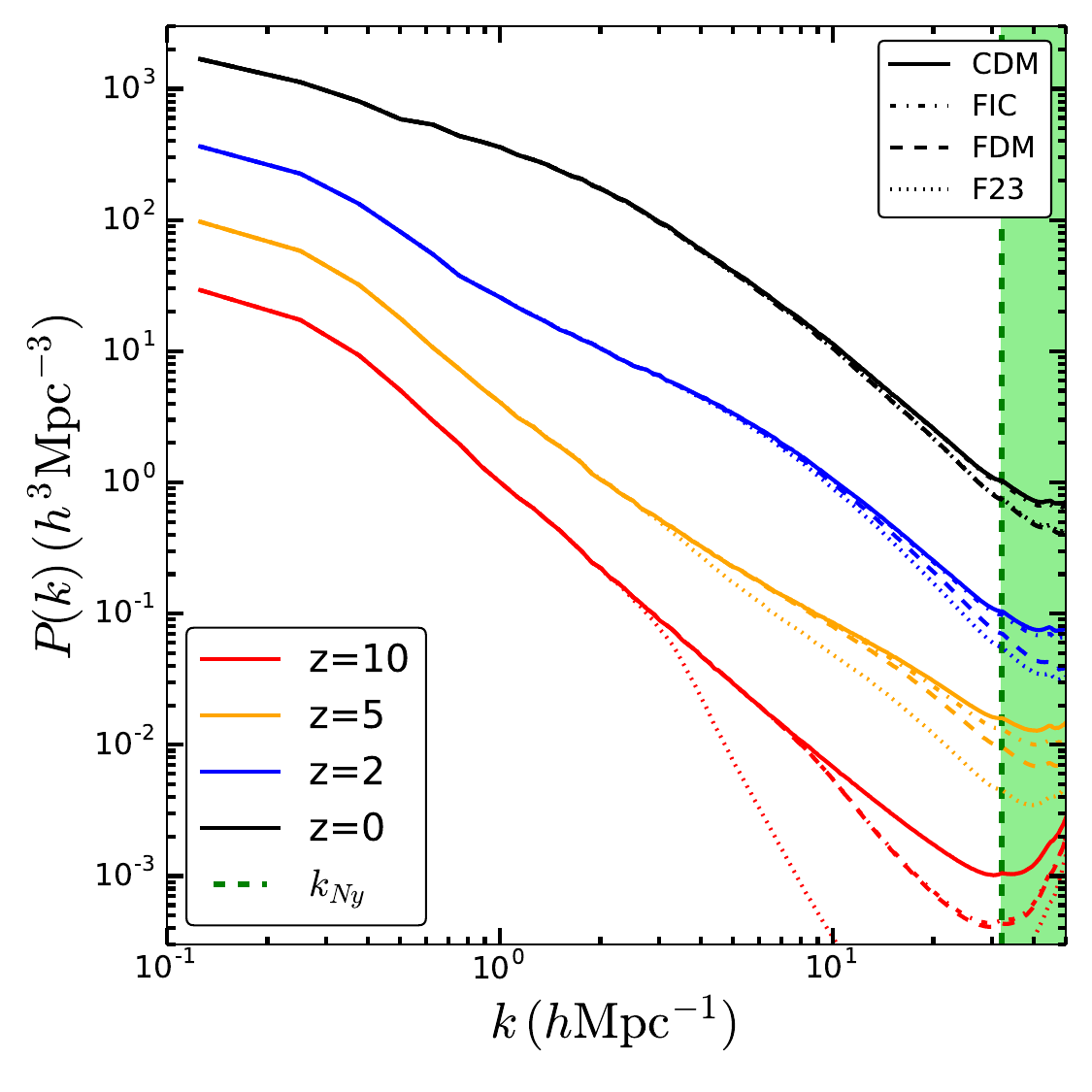}
\includegraphics[width=0.45\textwidth]{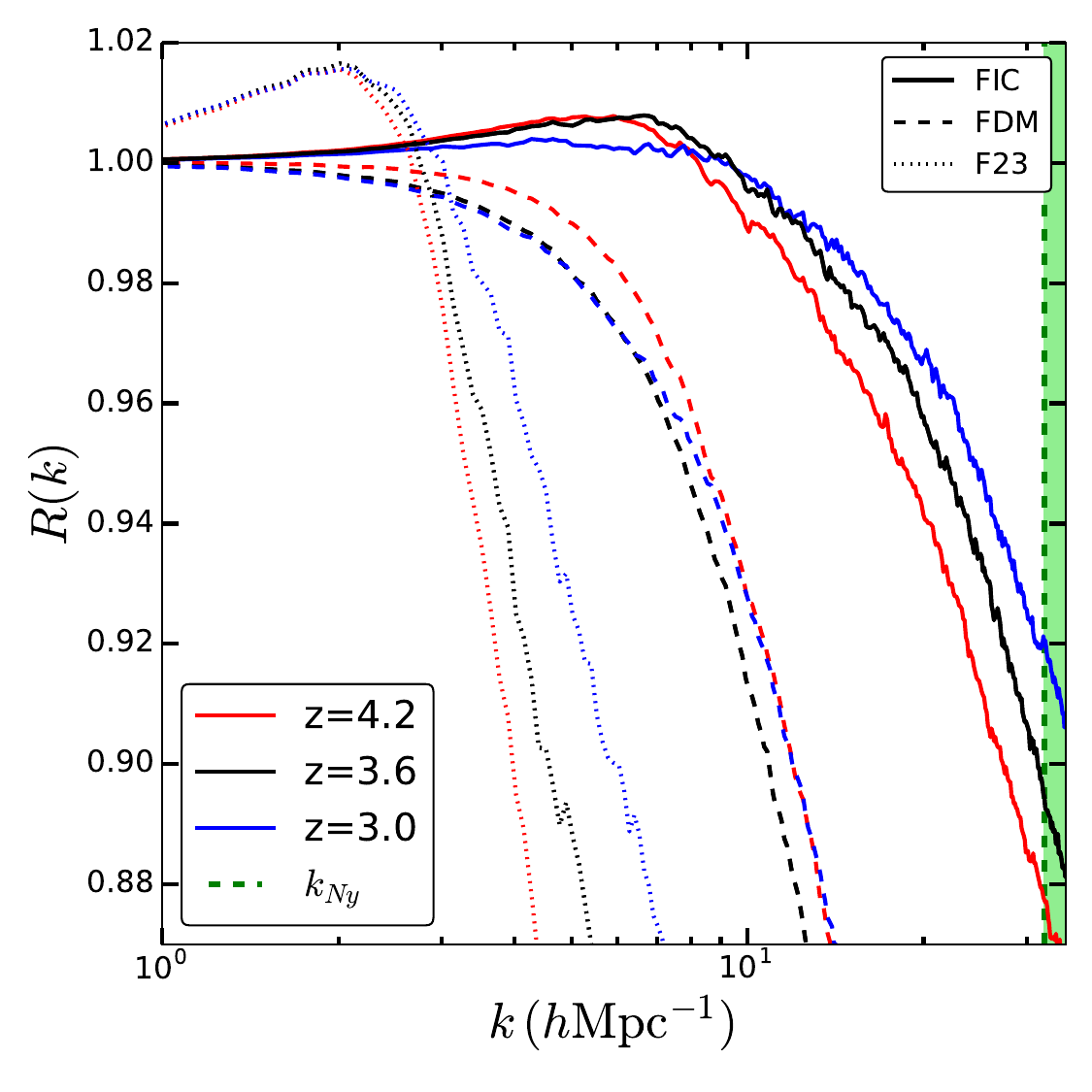}
\caption{
Left panel: the 3D power spectrum measured from 
the simulations \textbf{CDM} (in solid lines), 
\textbf{FIC} (in dot-dashed lines), 
\textbf{FDM} (in dashed lines) and 
an additional simulation \textbf{F23} with $m_\chi=2.5\times10^{-23}\ev$ (in dotted lines). 
Right panel: the relative non-linear matter power spectrum measured from the simulation. 
The ordinate axis $R(k)$ is the ratio of the 3D power spectra of the simulations \textbf{FDM} (in dashed lines), 
\textbf{FIC} (in solid lines) and \textbf{F23} (in dotted line) 
to that of the simulation \textbf{CDM}. 
Different colors represent different redshifts. 
The green dashed line represents the corresponding wavenumber of the Nyquist limit. 
}
\label{Fig:3Dpower}
\end{figure}


In Fig.~\ref{Fig:3Dpower}, we compare the 3D power spectra
 at different redshifts among 
different simulations \textbf{CDM}, \textbf{FIC}, \textbf{FDM} and \textbf{F23}.
In the left panel of Fig.~\ref{Fig:3Dpower}, 
the effect of FDM initial condition is almost negligible at low redshifts
but plays an important role at high redshifts. However, the effect of QP
is non-negligible even at low redshifts. 
During the non-linear gravitational evolution even without the QP, 
the power spectrum of the simulation \textbf{FIC}  
grows differently from the simulation \textbf{CDM}. 
However, at $z=0$ the power spectra of 
the simulations \textbf{FIC} and \textbf{CDM} almost overlap, and one can barely see the difference between them. 
The effect of the FDM initial condition turns out to be a tiny suppression 
on power spectrum at low redshifts.

On the other hand, the power spectrum of the simulation \textbf{FDM} is 
clearly different from the simulations \textbf{FIC} and \textbf{CDM} at low redshifts 
because of the non-trivial QP effect.
For the power spectrum of the simulation \textbf{F23}, 
the suppression on small scales is even more significant than 
all other simulations due to its larger QP.

In the right panel of Fig.~\ref{Fig:3Dpower}, 
we show the ratio $R(k)$ of the 3D power spectra from 
the simulations \textbf{FDM}, \textbf{FIC} and \textbf{F23} 
to that of simulation \textbf{CDM}. 
{
One can see that the QP suppresses the power spectrum by $2-5\%$
relative to the simulation \textbf{FIC} at $k<10\,h\,\mathrm{Mpc^{-1}}$ for three different redshifts $z=3.0,3.6,4.2$. 
However, for $k<1\,h\,\mathrm{Mpc^{-1}}$, the 3D power spectra from the simulations \textbf{FDM}, \textbf{FIC} and \textbf{F23} are identical. Therefore, we can conclude that, QP has no effect on large scale, which is well-expected. For $k>10\,h\,\mathrm{Mpc^{-1}}$, since it is approaching our resolution limit, the results should be taken carefully and critically.
}
%
%
The effect of QP in the simulation \textbf{F23} is clearly more significant 
than the simulation \textbf{FDM}.


\begin{figure}
\includegraphics[width=0.45\textwidth]{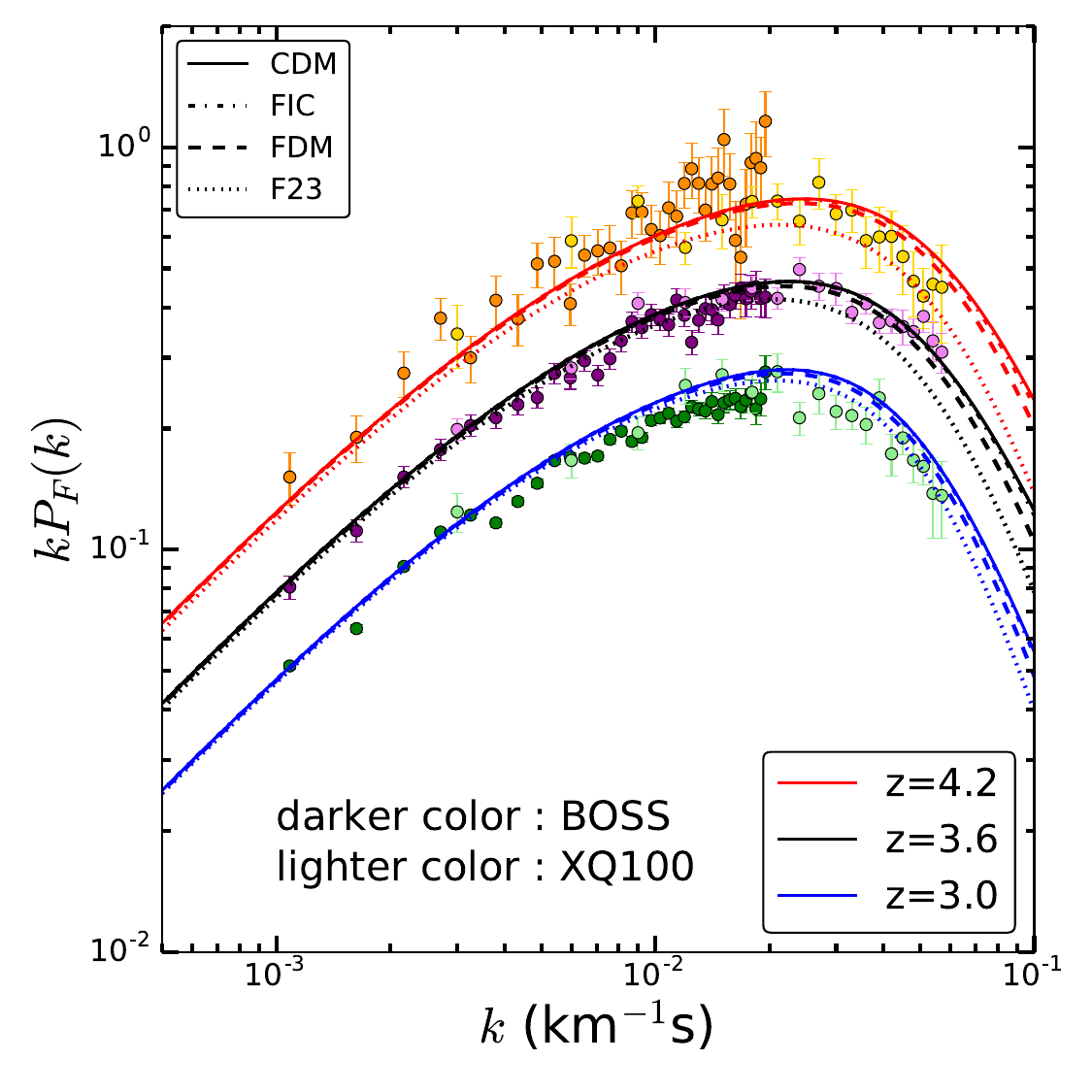}
\includegraphics[width=0.45\textwidth]{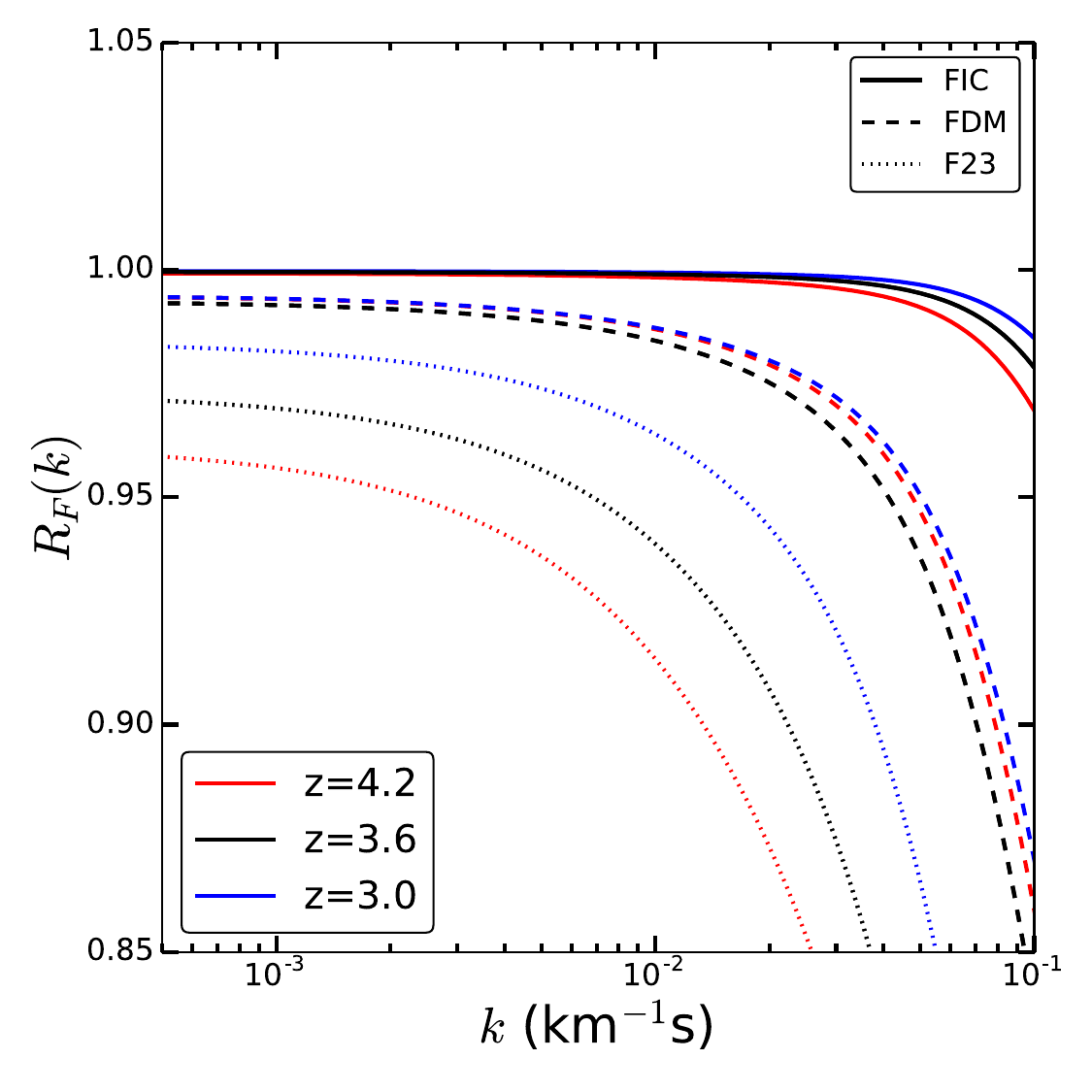}
\caption{Left panel: the 1D flux power spectra from the simulations \textbf{CDM} (in solid lines), \textbf{FIC} (in dot-dashed lines), \textbf{FDM} (in dashed lines) and \textbf{F23} (in dotted lines), with the data from BOSS (darker color) and XQ-100 (lighter color) at different redshifts (different color). The ordinate axis is the wavenumber $k$ times the 1D flux power spectrum and the abscissa axis is the corresponding wavenumber. Right panel: the impact of non-linear effect on the 1D flux power spectrum. The ordinate axis is the ratio of 1D flux power spectra of the simulations \textbf{FDM} (in dashed lines), \textbf{FIC} (in solid lines) and \textbf{F23} (in dotted line) to that of the simulation \textbf{CDM} and the abscissa axis is the corresponding wavenumber. The colors represent different redshifts.}
\label{Fig:lymandata}
\end{figure}

The 1D flux power spectra are shown in the 
left  panel of Fig.~\ref{Fig:lymandata} 
for comparison with the data of BOSS and XQ-100 at three different redshifts $z=3.0,3.6,4.2$. 
The solid, dot-dashed, dashed and dotted lines represent 1D flux power spectra from 
the simulations \textbf{CDM}, \textbf{FIC}, \textbf{FDM} and \textbf{F23}, respectively. 
The dots and error bars are the data from BOSS in darker colors and 
XQ-100 in lighter colors. 
{
At the small-scale region $10^{-2}\,\mathrm{km}^{-1}\,\text{s}<k<10^{-1}\,\mathrm{km}^{-1}\,\text{s}$,
which corresponds to $1\,h\,\mathrm{Mpc^{-1}}<k<10\,h\,\mathrm{Mpc^{-1}}$ in the discussion of the matter power spectrum,
the 1D flux power spectra for the simulations \textbf{FDM} and \textbf{F23} 
are relatively more suppressed than  those of \textbf{CDM} and \textbf{FIC}. 
Although the simulation \textbf{FIC} differs from \textbf{CDM} 
in the initial condition, the difference in their spectra is still small. 
}

The right panel of Fig.~\ref{Fig:lymandata} shows the ratio
$R_{F}(k)$ of the 1D flux power spectra of the simulations \textbf{FIC},
\textbf{FDM} and \textbf{F23} to that of the simulation \textbf{CDM}.
The degree of suppression is up to $10\%$ at
$k\simeq10^{-1}\,\mathrm{km^{-1}}\,\text{s}$ for the simulation \textbf{FDM}.
Additionally, the 1D flux power spectrum of the simulation
\textbf{F23} is suppressed even more than the simulation
\textbf{FDM}.  On the contrary, the degree of suppression is smaller
than $4\%$ at $k\simeq10^{-1}\,\mathrm{km^{-1}}\,\text{s}$ for the simulation
\textbf{FIC}.
For $k<10^{-2}\,\mathrm{km}^{-1}\,\text{s}$, the difference in the 1D flux power spectra for the simulations \textbf{FDM}, \textbf{FIC} and \textbf{CDM} is less than $2\%$.
In fact, the QP contribution cannot be neglected even if the QP is sub-dominant 
as shown in Fig.~\ref{Fig:quantumpressure}.

Finally, it is clear that a full investigation of 
hydrodynamic simulations with the QP is necessary
to robustly constrain the mass region of FDM. 


\section{Discussion}
\label{sec:discussion}
It has been reported in
Ref.~\citep{armengaud2017constraining,irvsivc2017first, 2017arXiv170800015K} that the FDM
mass can be excluded up to $10^{-21}\ev$ at $2\sigma$ significance.
However, the systematic uncertainties arising from 
both simulation (N-body and hydrodynamics) and 
gas properties (like the temperature of gas) were not discussed adequately in these works.
%

{
Note that the well-known discrepancies in simulation will introduce uncertainties on the observables, 
e.g. the code-to-code inconsistency in hydrodynamic simulations introduces at least $\sim 5\%$ uncertainty in 1D flux power spectrum as shown in Ref.~\citep{regan2007numerical,bird2013moving}. 
A brief discussion of the uncertainties in simulation is given in Appendix~\ref{Sec:uncertainty} for interested readers.
}

In the previous section, we consider the effects of FDM. However, the thermal properties of IGM also impact the 1D flux power spectrum, as shown in  Eq.~\eqref{eq:3Dto1D'} and Eq.~\eqref{eq:3Dto1D}. Since these two effects are degenerate in the linear power spectrum, 
one should consider them simultaneously before reaching any conclusion. 
In this subsection, we will discuss the possible uncertainty caused by 
the gas temperature and compare it with that cause by FDM mass to see 
their degeneracy in the power spectrum.


First, as an illustration of the potential effects of gas, we consider the prediction of the linear theory and its uncertainty. There are several parameters in Eq.~\eqref{eq:3Dto1D'}, but for simplicity we only consider the uncertainty of FDM particle mass $m_\chi$ and average gas temperature $T_0$. The error of the 1D flux power spectrum can be expressed as 
\be
	\label{eq:LinearUncertainty}
	\f{\Delta P_F}{P_F}=\f{\p\ln P_F}{\p m_\chi}\Delta m_\chi+\f{\p\ln P_F}{\p T_0}\Delta T_0.
\ee
We apply this formula to the linear 1D power spectrum at redshift $z=3.0$ and set $\gamma=1.64\left(1+3/5.5\right)^{-0.15}=1.72$, $\lambda_J=100\kpc$ in the calculation of partial derivatives.
For the most optimistic estimation,
we set $\Delta P_F=0$, so that we can express $\Delta m_\chi$ in terms of $\Delta T_0$.
In a most conservative precision $\Delta m_\chi/m_\chi<100\%$, 
one can obtain a maximum temperature uncertainty $\Delta T_{0,\max}$ and 
compare it to the current uncertainty of $T_0$. From Eq.~\eqref{eq:ThermalParameter}, the best-fit and 1$\sigma$ uncertainty of $T_0$ at redshift $z=3$ is about
\be
	T_0=20^{+7}_{-3}\times10^3\,\text{K}. \nonumber
\ee
For FDM mass $m_\chi=2.5\times10^{-23}\ev$, $\Delta T_{0,\max}$ is larger than $10^4\,\text{K}$. Therefore we can actually exclude this kind of FDM model robustly with Lyman-alpha observation. However, for FDM mass in the range from $10^{-22}\ev$ to $10^{-21}\ev$, we need much higher precision of the temperature to draw the conclusion (see Table.~\ref{Tab:TemperatureUncertainty}).

\begin{table}[!t]
\begin{center}
\begin{tabular}{p{3.5cm}<{\centering}|p{3.5cm}<{\centering}}
\hline
\hline
$m_\chi$ & $\Delta T_{0,\max}$\\
\hline
$2.5\times 10^{-22}\ev$ & $1.5\times10^3\,\text{K}$ \\ 
$5\times 10^{-22}\ev$ & $2\times10^2\,\text{K}$\\ 
$10^{-21}\ev$ & $1\times10^2\,\text{K}$\\ 
\hline
\hline
\end{tabular}
\caption{For each FDM mass $m_\chi$, the gas temperature uncertainty should be less than $\Delta T_{0,\max}$ for any confirmation or exclusion to be valid.}
\label{Tab:TemperatureUncertainty}
\end{center}
\end{table}

\begin{figure}\label{fig:degeneracy}
\includegraphics[width=0.45\textwidth]{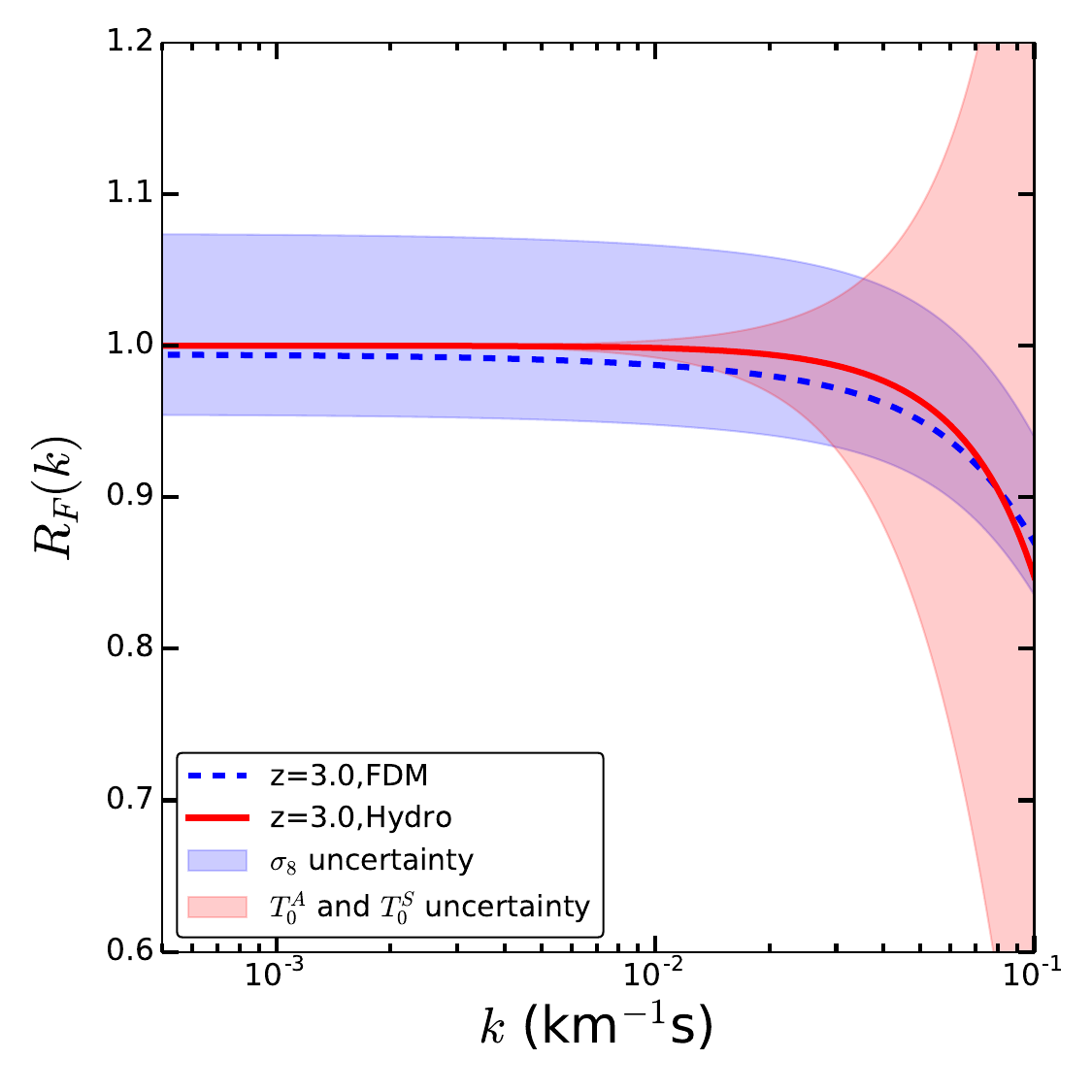}
\includegraphics[width=0.45\textwidth]{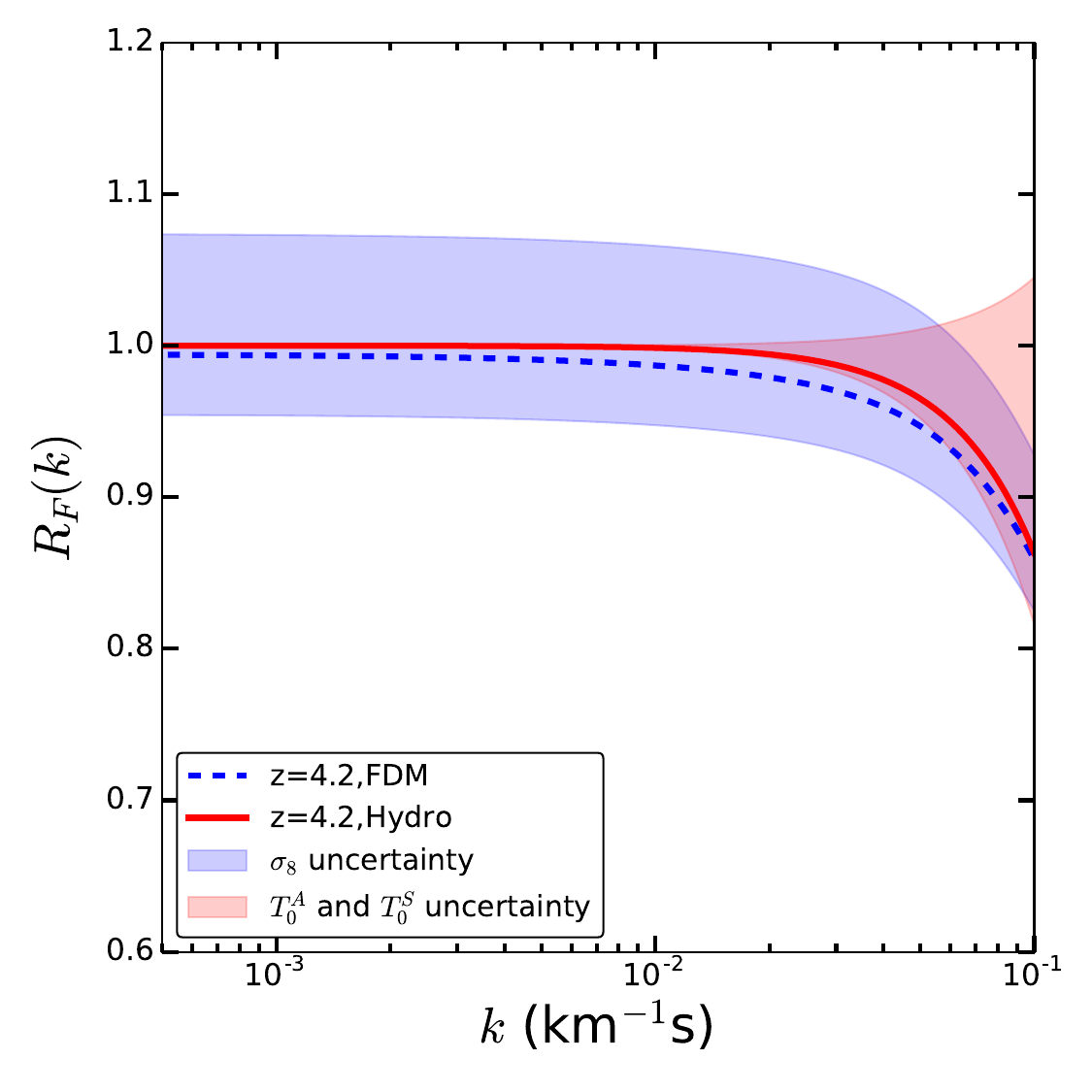}
\caption{Left (right) panel: the ratios of 1D flux power spectrum at $z=3.0$ ($z=4.2$). The blue dashed lines show the effect of FDM whose $m_\chi=2.5\times10^{-22}\,\mathrm{eV}$ measured from the simulations. The blue shaded area shows the 1$\sigma$ uncertainty range of $\sigma_8$ transferred to the normalization factor of the matter power spectrum. The red lines illustrate the effect of changing the temperature of the gas, which look similar to the blue dashed lines as an example. The red shaded area shows the 1$\sigma$ uncertainty range of $T_A$ and $T_S$. We can see that the suppression caused by FDM is fully contained within the uncertainty range of temperature, and so it is difficult to tell whether the suppression is due to dynamics or different temperatures of the gas.}
\label{Fig:ParameterDeg}
\end{figure}

Second, we use a similar method as in Sec.~\ref{subsec:ImpactLyalpha}, with one difference 
in the RHS of Eq.~\eqref{eq:DerivativePowerSp}; 
we multiply the ratio of bias function $W$ with different gas temperature 
rather than the ratio of matter power spectra,
\be
	P_{F,thermal}(k_z,z)=\int_{k_z}^\infty\f{k\dd k}{2\pi}P_{DM}(k,z)W(k)\f{W(k_z,k,T_0)}{W(k_z,k,T_{0,\text{bestfit}})}.
\ee
Here we used the Bias function in the linear theory Eq.~\eqref{eq:LinearBiasFunc}. 
Even though at the redshift range $z=3$ to $4$, the structure in the universe is non-linear,  we still use the ratio predicted by linear theory because our purpose here is to demonstrate the uncertainty of gas temperature instead of studying the thermal properties of gas which requires full hydrodynamics simulation. 

As in Sec.~\ref{subsec:ImpactLyalpha}, we will calculate the ratio $R_F=P_{F,thermal}/P_F$ again and compare it with the suppression induced by FDM. The best-fit value $T_{0,\text{bestfit}}(z)$ is calculated from Eq.~\eqref{eq:PowerLawParameterization} with $T_{0,\text{bestfit}}^A=9.2\times10^3\,\text{K}$ and $T_{0,\text{bestfit}}^S=-2.5$. The upper limit in the red shaded region in Fig.~\ref{Fig:ParameterDeg} is given by $T_{0}^A=9.1\times10^3\,\text{K}$ and $T_{0}^S=-2.05$ while the lower limit corresponds to $T_{0}^A=10.4\times10^3\,\text{K}$ and $T_{0}^S=-3$, which are chosen according to the 1$\sigma$ uncertainty Eq.~\eqref{eq:ThermalParameter}. From Fig.~\ref{Fig:ParameterDeg}, it is obvious that the current constraint on the temperature of gas is not good enough to exclude FDM mass $m_\chi=2.5\times10^{-22}\ev$, let alone the fact that we have not even included the uncertainties of other thermal parameters of the gas.
{
A detailed study of the thermal properties of the gas is needed to settle the issue.}

Finally, as Hui et al. pointed out in Ref.~\citep{Hui:2016ltb}, 
several additional astrophysical processes may also alter the 1D flux power spectrum, 
such as the effects of 
fluctuations in the ionizing background~\cite{Hui:2016ltb},
patchy reionization~\cite{Cen:2009bg}, 
modification of the thermal history~\cite{Garzilli:2015iwa},
and galactic outflows~\cite{Viel:2012sd}.
These additional factors are yet absent in many of the present cosmological simulations.
To sum up, the uncertainties from both numerical and physical factors prevent us from constraining the mass of FDM precisely, unless models and analyses with percent level precisions are available. 
A hydrodynamic simulation handling all of the uncertainties is 
needed in the future 
to set the correct constraint on the FDM mass.

\section{Conclusion}\label{Sec:conclusion}

In this paper, we have extended our FDM smoothed-particle hydrodynamics 
methodology to cosmological N-body simulation. 
Unlike previous works in literature, we have implemented not only 
the FDM initial condition 
but also the QP effect to our cosmological N-body simulations.  
The correct transformation of QP from physical to 
comoving coordinates has been derived in this work. 
With this new technique, we have performed four different simulations, 
\textbf{CDM}, \textbf{FIC}, \textbf{FDM}, and \textbf{F23}. 
We have shown the difference of over-density 
between \textbf{FIC} and \textbf{FDM} simulations. 
We have found that some granular structures located at 
higher density regions can be produced by QP. Remarkably, 
we are able to probe halos with mass smaller than 
$2\times10^{13}\,h\,{^{-1}M_{\odot}}$ in cosmological simulations 
based on our methodology.    

The matter power spectra from our four simulations tell us that the impact from QP 
is non-trivial,
as shown by the difference between \textbf{FIC} and \textbf{FDM}. 
Comparing with \textbf{CDM} in the region with wavenumber 
$k < 10h^{-1}$Mpc, the power spectrum suppression due to the effect of 
initial condition is less than $1\%$ at redshift $z=4.2$,
but the QP effect can cause $<5\%$ suppression in the same region.  
Hence, the impact from QP on the power spectrum is more 
significant than that from initial conditions at low redshifts. 
Moreover, the QP effect also depends on the redshift. At high
redshifts $z\sim 10$ the effect of modified initial condition is more important than 
QP, but vice versa at low redshifts $z\sim 0$.
Considering the DM mass around $10^{-23}\ev$, the matter power 
spectrum of the simulation \textbf{F23} 
shows a large deviation from that of \textbf{CDM} in the wavenumber 
$k\gtrsim 2 h^{-1}\Mpc $ region. 

Using the results of these four different simulations, 
we then further studied the flux power spectrum of Lyman-alpha forest.
We obtain the 1D flux power spectrum by integrating the 3D matter power spectrum 
taken from our simulation result. 
There is still suppression on the flux power spectra of \textbf{FDM} and \textbf{F23} on small scales compared to that of \textbf{CDM}.
However, the difference between \textbf{FIC} and \textbf{CDM} is small, which indicates that the suppression due to the transfer function of FDM is consumed by the non-linear evolution.
To summarize, by comparing the flux power spectra of different simulations, we demonstrate that the QP causes non-trivial effect on small scales, which could be important in the study of Lyman-alpha forest.

Finally, we discussed the uncertainties in simulations and models.  
{
As an example, we did a rough estimation and found that the average gas temperature $T_0$ and FDM mass $m_\chi$ are degenerate since a smaller $m_\chi$ or a higher $T_0$ both suppress the 1D flux power spectrum.}
We found that the current constraint on the average temperature of the gas is not 
accurate enough to exclude FDM mass $m_\chi >10^{-22}\ev$, 
which is summarized in Table.~\ref{Tab:TemperatureUncertainty} and Fig.~\ref{Fig:ParameterDeg}.

We conclude that the QP plays an important role in structure formation and affects the prediction for Lyman-alpha forest significantly. A further comprehensive hydrodynamic simulation including the QP and 
a precise constraint on the gas temperature of Lyman-alpha forest are needed 
to solidly set a lower bound on the FDM particle mass.

 

\section*{Acknowledgment}
We would like to acknowledge Eric Armengaud and Vid Ir\v{s}i\v{c} for discussion and Lachlan Lancaster for useful suggestions.
We also would like to acknowledge Lam Hui and Tom Broadhust for useful comments. This work is partially supported by a CUHK Discretionary Fund and by the MoST of Taiwan under the grant no.: 105-2112-M-007-028-MY3.


\appendix

\section{detail of the comoving transformation}
\label{app:comoving}
We show the full transformation of the QP from 
physical to comoving coordinates. 
The Lagrangian can be expressed as 
\begin{equation}\label{fulllagrangian}
L=\dfrac{1}{2}M\bm{v}^{\,2}-\dfrac{\hbar^2}{2{m_\chi}^2}\dfrac{M}{\bar{\rho}}(\nabla \sqrt{\rho})^2-M\phi(\bm{r}),
\end{equation}
where $M$, $v$, $m_\chi$, $\rho$, $\bar\rho$ and $\phi$ are mass and velocity of simulation particles, mass of FDM, density in physical coordinate, average density in physical coordinate and gravitational potential, respectively.
The first, second and the third terms in Eq.~\eqref{fulllagrangian} are the kinetic energy, 
the potential energy arising from the QP and the gravitational potential energy, respectively. 
We follow the basic transformations  
\begin{eqnarray}
\bm{r}=a\bm{x},\quad\bm{v}=\dot{\bm{r}}&=&a\dot{\bm{x}}+\dot{a}\bm{x},\quad\rho=\rho_x/a^3,\nonumber\\
\quad\nabla_x=a\nabla_r ,\quad\Delta_x&=&a^2\nabla_r\cdot\nabla_r=a^2\Delta_r,
\end{eqnarray}
which incorporate the scale factor $a$ to account for the expansion of the universe. 
Here $r$ denotes the physical coordinates and $x$ denotes the comoving 
coordinates. 

After including the transformations, the Lagrangian Eq.~\eqref{fulllagrangian} becomes
\begin{equation}
L=\dfrac{1}{2}M(a\dot{\bm{x}}+\dot{a}\bm{x})^2-\dfrac{1}{a^2}K_{\rho x}-M\phi(\bm{x}),
\end{equation}
where we define
\begin{equation}
K_{\rho x}=\dfrac{\hbar^2}{2m_\chi^2}\dfrac{M}{\bar{\rho_x}}(\nabla_x\sqrt{\rho_x})^2.
\label{eq:krhox}
\end{equation}
The transformation of the Poisson equation can be written as  
\begin{equation}
\Delta_r\phi=4\pi G(\rho-\rho_{\Lambda}) \to
\Delta_x\phi=\dfrac{4\pi G}{a}(\rho_x-\rho_{\Lambda x}),
\end{equation}
where $\rho_\Lambda$ is the density of the cosmological constant.
In order to simplify the equation, we perform a canonical transformation on the Lagrangian
\begin{equation}\label{canonicaltransformation}
\begin{gathered}
L=L-\dfrac{dF(\bm{x},t)}{dt}, \\
F=\dfrac{1}{2}Ma\dot{a}\bm{x}^{\,2}, \\
\end{gathered}
\end{equation}
which does not change the equation of motion. 
Now the Lagrangian can be written as
\begin{equation}
L=\dfrac{1}{2}Ma^2\dot{\bm{x}}^{\,2}-\dfrac{1}{a^2}K_{\rho x}-M\Phi,
\end{equation}
where we define $\Phi=\phi+(1/2)a\ddot{a}\bm{x}^2$ for simplicity.
Consequently, the Poisson equation is converted to
\begin{equation}
\Delta_x\Phi=\Delta_x(\phi+\dfrac{1}{2}a\ddot{a}\bm{x}^{\,2})=\dfrac{4\pi G}{a}(\rho_x-\rho_{\Lambda x})+3a\ddot{a}.
\end{equation}
From the second equation of the Friedmann equations,
\begin{equation}\label{secondfriedmann}
\dfrac{\ddot{a}}{a}=-\dfrac{4\pi G}{3}(\bar{\rho}-\rho_{\Lambda})=-\dfrac{4\pi G}{3a^3}(\bar{\rho_x}-\rho_{\Lambda x}),
\end{equation}
one can obtain
\begin{equation}
\Delta_x\Phi=\dfrac{4\pi G}{a}(\rho_x-\bar{\rho_x}).
\end{equation}
Additionally, we define $\Psi=a\Phi$ and acquire $\Delta_x\Psi=4\pi G(\rho_x-\bar{\rho_x})$.
Now the Lagrangian can be written as 
\begin{equation}
L=\dfrac{1}{2}Ma^2\dot{\bm{x}}^{\,2}-\dfrac{1}{a^2}K_{\rho x}-\dfrac{1}{a}M\Psi.
\end{equation}
Hence we can define the canonical momentum as 
\begin{equation}\label{canonicalmomentum}
\bm{p}=\dfrac{\partial L}{\partial\dot{\bm{x}}}=Ma^2\dot{\bm{x}},
\end{equation} 
and write down the Hamiltonian
\begin{equation}\label{comovinghamiltonian}
H=\dfrac{1}{2Ma^2}\bm{p}^{\,2}+\dfrac{1}{a^2}K_{\rho x}+\dfrac{1}{a}M\Psi.
\end{equation}
As a consequence, we obtain the equations of motion
\begin{equation}\label{comovingEOM}
\begin{gathered}
\dot{\bm{x}}=\dfrac{\partial H}{\partial \bm{p}}=\dfrac{\bm{p}}{Ma^2}, \\
\dot{\bm{p}}=-\dfrac{\partial H}{\partial \bm{x}}=-\dfrac{1}{a^2}\nabla_x K_{\rho x}-\dfrac{M}{a}\nabla_x\Psi.
\end{gathered}
\end{equation}
In the equations of motion, the terms for gravity and QP have prefactors $1/a$ and $1/a^2$ respectively, and hence we need to treat them separately in the comoving coordinates. In the physical coordinates, 
we can deal with these two terms simultaneously since $a=1$.

\section{simulation uncertainty}\label{Sec:uncertainty}

The systematic uncertainties in N-body simulation are well studied. The matter power spectrum is affected by the methodology of generating initial condition for simulations, such as the choice between the first order Lagrangian perturbation theory (1LPT) and the second order Lagrangian perturbation theory (2LPT) can introduces $\sim 6\%$ difference in the matter power spectrum, as reported in Ref.~\citep{l2014effects}. The halo mass function in the large mass end is sensitive at $\sim 7\%$ level to the choice of the finite simulation box size, which can be corrected owing to their clear nature~\citep{bagla2006effects}.

The uncertainty resulting from different N-body simulation codes 
is numerically hard to estimate, and the origin of the discrepancies among
different codes is also barely known. 
Such kind of issues are comprehensively discussed in Ref.~\citep{heitmann2010coyote}.
Roughly speaking, the discrepancy between different codes is within $10\%$ 
in the matter power spectrum and halo mass function. 
However, Ref.~\citep{kim2013agora,sembolini2016nifty} reported 
a $20\%$ discrepancy in the center and $10\%$ in the outskirts of the halo 
by comparing the halo density profile from several different codes. 


Unlike the $\mathcal{O}(10\%)$ errors in the N-body simulations, 
the uncertainties involved in hydrodynamic simulations are much larger. 
Usually, a considerable uncertainty can be introduced due to 
the treatments of the gaseous component. 
As demonstrated in Ref.~\citep{o2005comparing,vazza2011comparison,kim2013agora,sembolini2016nifty}, 
a large inconsistency of the galaxy structure can be caused by using different hydrodynamic codes.  
Quantitatively, the errors of the gas density and 
temperature in the centers of the galaxies are about one to two orders of magnitude, estimated  by comparing 
different hydrodynamic codes.


The one-dimension (1D) flux power spectrum of Lyman alpha forest 
can also contain some level of uncertainties, 
e.g., see the code comparison in Ref.~\citep{regan2007numerical,bird2013moving}.
For hydrodynamic simulations using \texttt{ENZO}~\citep{bryan2014enzo} and using \texttt{Gadget2}, 
there is also a $\sim 5\%$ difference in the 1D flux power spectrum and $\sim 10\%$ 
difference in the probability distribution of density and temperature.
Moreover, for simulations using \texttt{AREPO}~\citep{springel2010moving} and \texttt{Gadget2}, 
there are also $5\%$ difference in their 1D flux power spectra.

\bibliographystyle{h-physrev}
\bibliography{MDM}

\end{document}